\documentclass[sigconf]{acmart}

\usepackage{blindtext}

\AtBeginDocument{%
  \providecommand\BibTeX{{%
    \normalfont B\kern-0.5em{\scshape i\kern-0.25em b}\kern-0.8em\TeX}}}

\copyrightyear{2020}
\acmYear{2020}
\setcopyright{acmlicensed}\acmConference[OzCHI '20]{32ND AUSTRALIAN
CONFERENCE ON HUMAN-COMPUTER INTERACTION}{December 2--4,
2020}{Sydney, NSW, Australia}
\acmBooktitle{32ND AUSTRALIAN CONFERENCE ON HUMAN-COMPUTER
INTERACTION (OzCHI '20), December 2--4, 2020, Sydney, NSW, Australia}
\acmPrice{15.00}
\acmDOI{10.1145/3441000.3441031}
\acmISBN{978-1-4503-8975-4/20/12}

\begin{document}

\title{A Tangible Multi-Display Toolkit to Support the Collaborative Design Exploration of AV-Pedestrian Interfaces}


\author{Marius Hoggenmueller}
\email{marius.hoggenmueller@sydney.edu.au}
\affiliation{Design Lab, Sydney School of Architecture, Design and Planning
  \institution{The University of Sydney}
}

\author{Martin Tomitsch}
\email{martin.tomitsch@sydney.edu.au}
\affiliation{Design Lab, Sydney School of Architecture, Design and Planning
  \institution{The University of Sydney}
}
\affiliation{CAFA Beijing Visual Art Innovation Institute, China}

\author{Callum Parker}
\email{callum.parker@sydney.edu.au}
\affiliation{Design Lab, Sydney School of Architecture, Design and Planning
  \institution{The University of Sydney}
}

\author{Trung Thanh Nguyen}
\email{thanh.trung.nguyen@sydney.edu.au}
\affiliation{Design Lab, Sydney School of Architecture, Design and Planning
  \institution{The University of Sydney}
}

\author{Dawei Zhou}
\email{dawei.zhou@sydney.edu.au}
\affiliation{Design Lab, Sydney School of Architecture, Design and Planning
  \institution{The University of Sydney}
}

\author{Stewart Worrall}
\email{stewart.worrall@sydney.edu.au}
\affiliation{Australian Centre for Field Robotics
  \institution{The University of Sydney}
}

\author{Eduardo Nebot}
\email{eduardo.nebot@sydney.edu.au}
\affiliation{Australian Centre for Field Robotics
  \institution{The University of Sydney}
}


\renewcommand{\shortauthors}{Hoggenmueller et al.}

\begin{abstract}
The advent of cyber-physical systems, such as robots and autonomous vehicles (AVs), brings new opportunities and challenges for the domain of interaction design. Though there is consensus about the value of human-centred development, there is a lack of documented tailored methods and tools for involving multiple stakeholders in design exploration processes. In this paper we present a novel approach using a \textit{tangible multi-display toolkit}. Orchestrating computer-generated imagery across multiple displays, the toolkit enables multiple viewing angles and perspectives to be captured simultaneously (e.g. top-view, first-person pedestrian view). Participants are able to directly interact with the simulated environment through tangible objects. At the same time, the objects physically simulate the interface's behaviour (e.g. through an integrated LED display). We evaluated the toolkit in design sessions with experts to collect feedback and input on the design of an AV-pedestrian interface. The paper reports on how the combination of tangible objects and multiple displays supports collaborative design explorations. 

\end{abstract}



\begin{teaserfigure}
  \includegraphics[width=\textwidth]{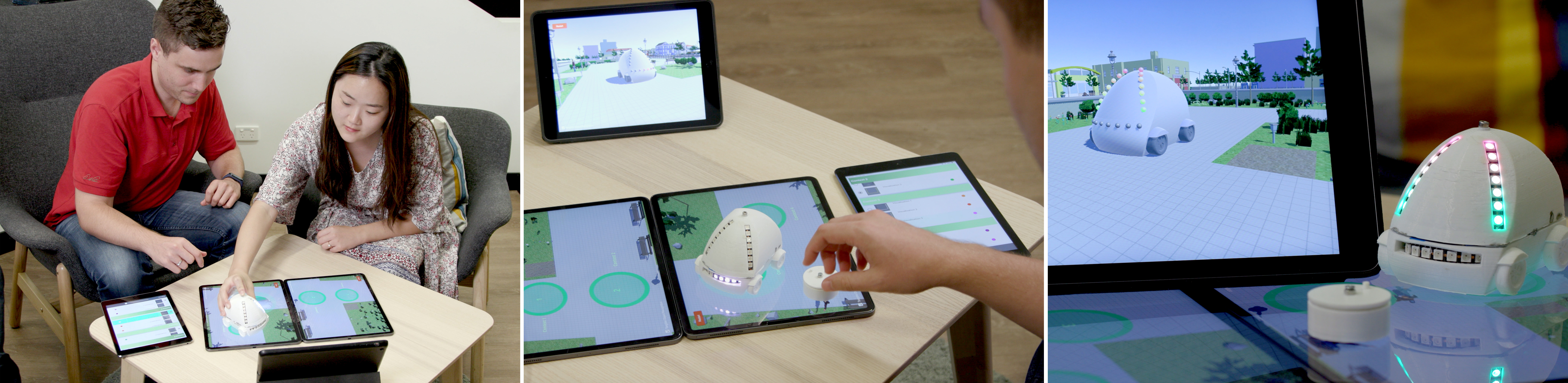}
  \caption{We present a tangible multi-display toolkit to inform the design of AV-pedestrian interfaces. The toolkit contains a) computer simulations across multiple displays to capture different viewing angles, b) tangible objects to interact with the simulated environment and to simulate the interface's behaviour through an integrated LED display, and c) a configuration app allowing participants to change the interface's behaviour in real-time.}
  \Description{Users interacting with the tangible multi-display toolkit.}
  \label{fig:teaser}
\end{teaserfigure}

\maketitle


\section{Introduction}
The rapid development in computing, sensing and network technologies has enabled the creation of systems, where digital and physical processes are increasingly intertwined and automated, able to communicate with each other, and interactions are highly context dependent \cite{Garro2019}. This is the realm of cyber-physical systems and environments \cite{Forshaw2012} ranging from applications in the domestic context \cite{Nazli2017}, or at a larger scale, in the smart city. Here, we can see now an increasing deployment of mobile service robots and the advent of autonomous vehicles (AVs) in the near future \cite{Macrorie2019}. The emergence of such highly automated cyber-physical systems also brings new opportunities and challenges to the domain of interaction design \cite{Pettersson2017}. For example, autonomous driving will enable a multitude of non-driving-related activities and interactions with devices, services and other passengers, that are still yet to be defined \cite{Pfleging2016, Schroeter2012}. Further, with the rise of automation and system agency, there is a pressing need for intuitive human-machine communication to keep the user in the loop \cite{Kun2016}. In the context of AVs, research has for example investigated the use of in-vehicle user interfaces to inform passengers about the system's intent \cite{Sanders2010}; further, there is an increasing stream of research focusing on communication with other road users \cite{Nguyen2019, Owensby2018, Mahadevan2018}, which at the same time also underlines the complexity of the design context as a wide range of ``user'' perspectives needs to be considered \cite{Boll2019}.\newline

As a result, researchers have advocated that new methods and tools are required for designing future interactions with cyber-physical systems \cite{Pettersson2017, Tomitsch2020, Malizia2018}. Such methods and tools should aid designers to manifest early visions in the form of (semi-)functional prototypes \cite{Schmidt2017}, which in turn can ease the involvement of stakeholders during the design process \cite{Wiethoff2012, Wiethoff2013}. Participatory and collaborative approaches are in particular relevant in the AV context, as the systems will operate in real-world urban environments with far-reaching implications on humans. This requires the viewpoints and expertise of various stakeholders, including engineers, urban planners and citizen. However, to date the availability and systematic documentation of collaborative approaches and purpose-built tools remains rare. Traditional methods and tools, such as paper prototypes and 3D mock-ups for scale scenarios \cite{Pettersson2017}, have limitations when it comes to representing cyber-physical systems and environments that blend digital and physical user interfaces. Further, the ability to simulate the dynamic and complex nature of situations and interactions that occur between people, system and environment is limited. 
In recent years, immersive virtual reality (VR) has been successfully used for the risk-free and contextualised evaluation of interactions with AVs \cite{Nascimento2019, Stadler2019, Flohr2020, Nguyen2019}. However, the 
complexity of building VR environments is not well-suited to support early design explorations and the individualised experience of VR limits the value of this approach for collaborative design explorations.\newline

To address this gap, we developed a toolkit approach that uses a combination of commercially available tablets and custom-built tangible objects (see Fig. \ref{fig:teaser}). The study was embedded in a larger research project involving a real ride-sharing AV with a low-resolution lighting display to communicate the AV's intent and awareness to nearby pedestrians and prospective users (in our case, riders). In this paper, we report on the use of the toolkit to evaluate a series of AV-pedestrian interface designs through design sessions with experts. Comparing data from sessions involving the toolkit with sessions carried out with a simple first-person simulation view, we discuss the value of the tangible multi-display toolkit for a) understanding the context, b) facilitating communication, followed by a report on the usability of our proposed toolkit. We further reflect on the toolkit approach and its broader value for evaluating cyber-physical systems and their interfaces. The contributions of the paper are: First, a generalisable toolkit approach for engaging experts in collaboratively evaluating early interface designs for AVs and other cyber-physical systems. Second, insights on how the combination of tangible objects and multiple displays supports collaborative design exploration. Third, a reproducible description of the toolkit and its setup along with the software components and 3D models of the tangibles used in our study. 



\section{Related Work}

This paper builds on and contributes to two areas of related work: Through the case study that provided the context for the toolkit development and evaluation, we contribute to AV-pedestrian interfaces; through the toolkit itself we contribute to the larger field of prototyping and design tools. Below we highlight relevant previous work that served as a foundation for our study across those two areas. 

\subsection{AV-Pedestrian Interfaces}
In recent years, researchers have stressed that building trust in autonomous vehicles is one of the key challenge to ensure that this technology will successfully be deployed on the roads and finds wide acceptance in society \cite{Lee2018}. Therefore, amongst others AVs need to communicate their status, intent and awareness, not only to people inside the car, but also to other road users in the urban environment, such as pedestrians \cite{Owensby2018}. As a consequence, researchers have investigated the use of a wide range of external human-machine interfaces (eHMI) \cite{Clercq2019} including projection-based solutions \cite{Nguyen2019}, or displays attached to various positions of the vehicle \cite{Eisma2020} and supporting various communication modalities (e.g. light band eHMIs for abstract representations \cite{Debargha2020}, or higher resolution displays for text and symbols \cite{Hollaender2019}). Most of these interfaces have been validated through experimental studies in VR with potential users \cite{Stadler2019}. Further, a systematic review on eHMIs has emphasised that a majority of these studies are limited to one specific traffic scenario, mostly focusing on an uncontrolled zebra crossing scenario \cite{Dey2020}. This opens up questions how to design and evaluate a more comprehensive set of communicative means, which are consistent within a wide range of scenarios and taking into account the various contexts, in which AVs will be operating in (e.g. streets, shared spaces).

\subsection{Prototyping and Design Tools}
The creation of prototypes -- ranging from simple paper representations \cite{Wiethoff2013} to fully functional mock-ups \cite{Schmidt2017} -- is an integral part of the interaction design process \cite{Buxton2007}. During a product development cycle, prototypes can fulfil various purposes for the design team: Often prototypes are being used to test and evaluate certain aspects of a design before moving on to the next development stage. Lim et al. highlight the importance of understanding prototypes as filters to manifest different qualities of the envisioned product through the careful selection of prototyping materials and resolution \cite{Lim2008}. Besides testing and evaluation, prototypes are also being used in a more generative manner, for example, to foster ideation processes and to facilitate communication among various stakeholders, thereby considering prototypes as ``tangible thinking tools'' \cite{Lockton2020, Rygh2019}. In this vein, Henderson et al. for example report how they employed rapid modular prototypes to augment user interviews in the context of a study on parking meters, leading to richer discussions between users and design researchers \cite{Henderson2018}.

As the creation of prototypes can be cumbersome and time consuming, the use of modular prototypes and making approaches replicable and adaptable through prototyping toolkits is considered as an important contribution in HCI \cite{Ledo2018}. In the automotive context, researchers for example published prototyping approaches and toolkits for designing novel in-vehicle user interfaces \cite{Lauber2014, Broy2014, Broy2016, Gerber2019}. However, these toolkits are mainly focusing on the interface qualities themselves, rather then simulating interaction effects between people, system and environment at large. However, in particular in the context of AV-pedestrian interfaces an understanding of the broader context is important \cite{Kray2007}. Therefore, researchers have argued for the importance of context-based prototyping techniques, which can simulate dynamic and complex situations occurring around cyber-physical systems that are deployed in real-world (urban) environments \cite{Flohr2020}. Further, researchers have highlighted the need for tools and techniques to make these situations more tangible \cite{Pettersson2017} and inclusive for the involvement of a wide range of stakeholders in early design exploration sessions \cite{Malizia2018}.

\section{Research Context} 
The conceptualisation and development of the prototyping toolkit presented in this paper is embedded within the context of a larger research project on shared autonomous mobility. Specifically, the study for which we developed the toolkit focused on the design of an AV-pedestrian interface, which would indicate to the rider which car is picking them up while at the same time informing pedestrians about the AV's status and intent. The AV-pedestrian interface was conceptualised for a real-world AV, involving researchers from robotic engineering, urban planning, social science and interaction design.\newline 

\begin{figure}[b]
  \includegraphics[width=\linewidth]{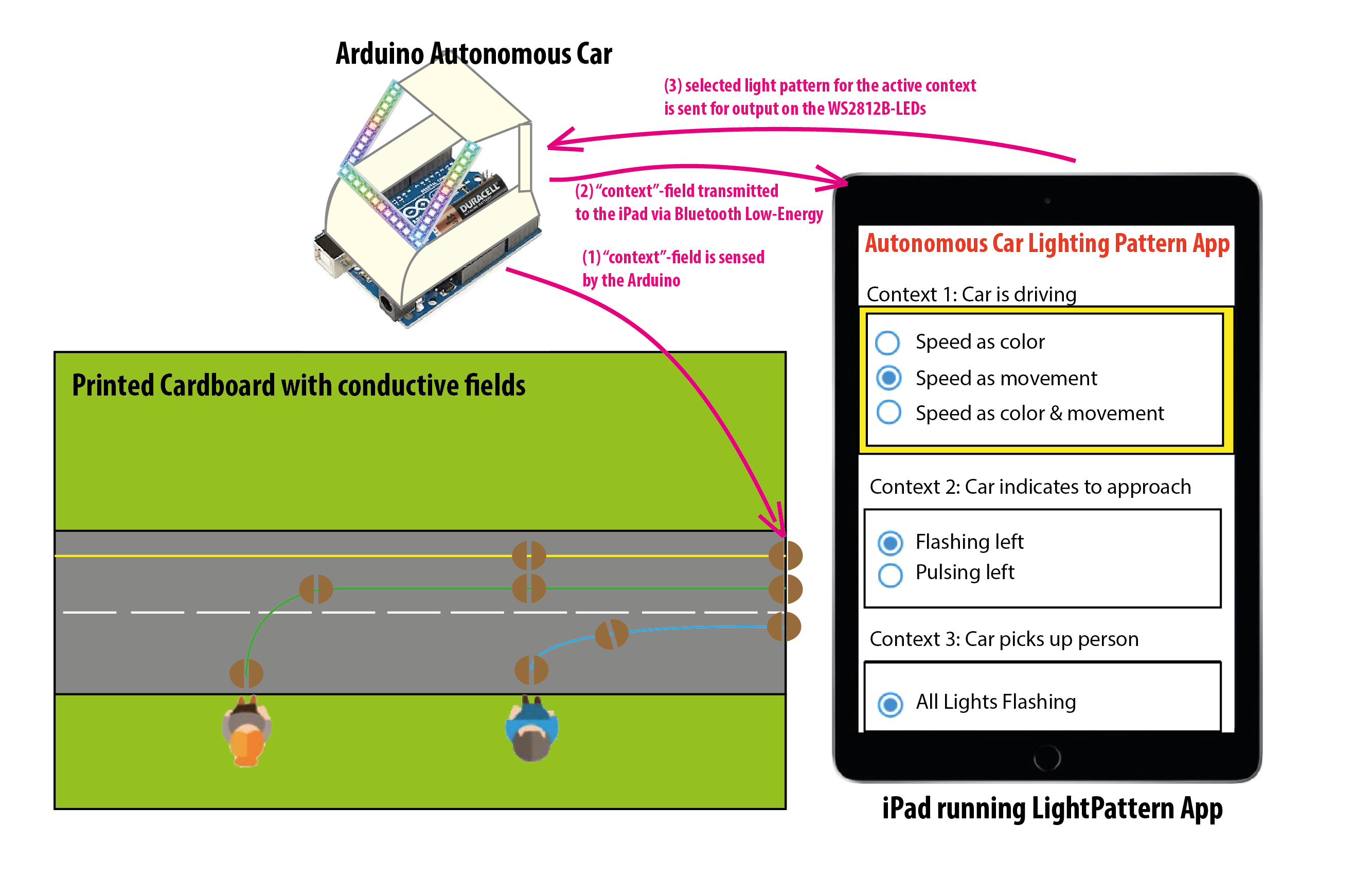}
  \caption{Early concept of the prototyping toolkit to evaluate various light patterns.}
  \Description{Early concept of the prototyping toolkit.}
  \label{fig:early_prototype}
\end{figure}

Early on in the project we decided on a low-resolution (low-res) lighting display due to the flexibility of LEDs, allowing them to be integrated into various shapes while offering high contrast and hence better visibility. Low-res lighting displays have further been used in previous HCI research studies to communicate information through simple visual cues (i.e. colours, temporal change of visual elements) which can be also perceived from a distance \cite{Hoggenmueller2015}. In our project team meetings we discussed various design concepts for different situations and scenarios, using video mock-ups created in Adobe After Effects as early representations of the low-res display. We quickly realised the limited efficacy of this approach, as we would need to render out new videos for every iteration, to share them with the team. Further, we were confronted with the issue that we would not be able to evaluate how the light patterns would look like with the actual hardware before implementation, which has been previously identified as a challenge when designing low-res lighting displays \cite{Wiethoff2010}. As a consequence, the interaction design team developed an early concept for a prototyping toolkit consisting of an Arduino, wrapped around a car paper prototype with LEDs, which would be controlled by a configuration app running on an iPad to quickly adapt and program light patterns for various situations (see Figure \ref{fig:early_prototype}). The intention was to allow quicker iterations and to enable the interaction design team to present design concepts during team meetings in a more tangible form and at a higher fidelity level \cite{Lim2008}.\newline

As we started developing this concept further, we also faced an increasing number of questions and challenges regarding the design of the light patterns. Even though there is an emerging body of previous work to build on \cite{Debargha2020, Benderius2018, Mahadevan2018}, most of that work provides design recommendations for a particular situation and modality (e.g. which colours to use for traffic negotiations with pedestrians at intersections \cite{Debargha2020}). As we were developing light patterns for an actual AV, we needed to cater for a more comprehensive range of scenarios. 
Also falling back on general guidelines for ambient light systems \cite{Matviienko2015, Harrison2012} that are based on people's existing associations and therefore applicable to a wide range of systems and products, can fall short in the more complex context of AVs. For example, previous research has highlighted the confusion with a red light signal on a moving vehicle, as it could either refer to the car's internal state, which would mean that the car is stopping/stationary, or implement a traffic light metaphor, which would signal other road users to stop \cite{Rouchitsas2019}. In light of these considerations, we concluded that before implementing the light patterns on the actual AV, we needed to find a way to test the AV-interface with experts that could offer various perspectives and ensure that the final design would satisfy the users' expectations. 
In order to support collaborative design explorations through our toolkit approach, we refined our initial concept and deduced additional requirements, such as supporting development and evaluation cycles, allowing for quick adaptations of the interface elements, and incorporating a higher level of fidelity so that also stakeholders outside the automotive community could imagine and relate to the simulated environment.


\begin{figure*}[t]
  \includegraphics[width=1\textwidth]{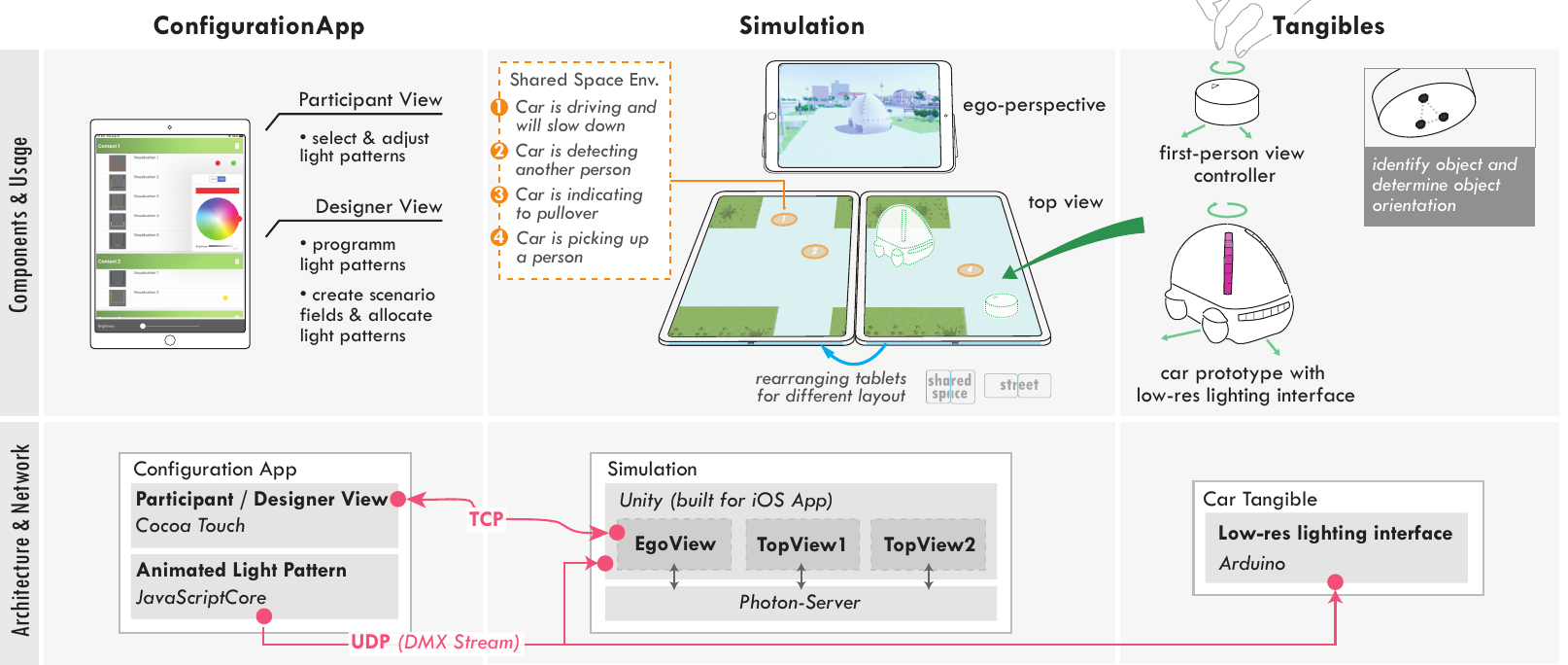}
  \caption{Overview of the various components of the toolkit and its usage (top), implementation details and network communication between the different components (bottom).}
  \Description{Overview of the toolkit components.}
  \label{fig:toolkit_overview}
\end{figure*}

\section{Toolkit}
In the following section we present the final design of our \textit{tangible multi-display toolkit}, which consists of three core components (see Figure \ref{fig:toolkit_overview}): a) 3D computer-generated simulations displayed on various tablets, b) tangible objects to control the simulation, and c) a configuration app running on a tablet to control and adjust the design options. The toolkit was designed to be capable of representing different \emph{environments} and a series of \emph{scenarios} within the environment. For example, one of our environments represented a shared urban space. The scenarios within that environment included, for example, the vehicle turning around a corner, approaching a pedestrian and coming to a halt.

The toolkit and its components where customised to support prototyping the low-res AV-pedestrian interface for our real AV. However, we expect that our toolkit approach can be used for prototyping a wide range of AV-pedestrian interfaces \cite{Flohr2020} and other types of cyber-physical systems (such as urban robots) more broadly. In this section we present the different components and their usage, followed by the implementation details. All software components and the 3D models of the tangibles are available via github\footnote{\url{http://ryanntt.com/tangible-multi-display-toolkit/}} to allow replication and further development by others.

\subsection{Components and Usage}
The 3D computer-generated \emph{simulation} running on three portable tablet devices provides the foundation for the interactive and animated scale scenario explorations. In our particular setup, we used two Apple iPad Pro (12.9 inches) tablets placed next to each other horizontally on a tabletop. These two tablets simulated an environment from the top-view perspective, with the environment spanning across both tablets. An Apple iPad Air (10.5 inches) tablet placed at the top end of the two top-view tablets and supported by a cover stand displayed the same simulated environment from a first-person pedestrian view. We designed two different environments: a shared space environment with an AV driving on a public plaza, for which the tablets were aligned on the long edges roughly forming a square (see active configuration in Figure \ref{fig:toolkit_overview}), and a street environment, for which the tablets were rearranged and aligned on the short edges.

For interaction with the simulated environment, we created two \emph{tangibles} that can be placed on the top-view tablets: A first-person view controller coming in a cylindrical shape with an engraved arrow to indicate the viewing direction, and a car prototype with an integrated low-res lighting display. Depending on how the first-person view controller is placed on the top-view tablet, the camera position and viewing angle of the first-person pedestrian view, which simulates a waiting rider, change accordingly in real time. The car prototype can be placed on various predefined positions (so-called \emph{scenario fields}), which are highlighted through displayed overlays on the top view. When the car prototype is placed on one of those positions, a short animated sequence of this particular scene (e.g. AV turning to the left) is displayed in a loop on the first-person view. The physical car prototype and the animated vehicle in the simulation have the same shape to clearly indicate that the physical prototype is a representation of the simulated vehicle. Both feature a ``U''-shaped low-res display with 21 single-controllable lights in total.

Further, our toolkit comes with a \emph{configuration app} for two main purposes: First, to allow participants to select and adjust the low-res light patterns during collaborative design explorations, and second, to support designers in developing light patterns and making adjustments to the setup itself. To address both purposes, we designed a participant and designer view. The \emph{participant view} shows a list of the different \emph{scenario fields} and various pre-configured light patterns which can be selected for a particular scenario (e.g. ``light band sweeping to the left'' light pattern for the ``AV turning left'' scenario). In the list view of the configuration app, the currently active scenario (determined by the position of the physical car prototype) is highlighted. When selecting a light pattern from the list, the animation is streamed to both the low-res display on the physical and the simulated AV prototype. Further, each light pattern can be modified in terms of colours being used. For example, if a light pattern is composed of two different colours, two colour fields are visible and can be modified via a colour picker pop-up. Further, in the participant view, the overall brightness of the light patterns can be controlled through a slider. 

The \emph{designer view} allows for more sophisticated options for configuring the toolkit. An important and unique feature of the toolkit is that it enables designers to create and program new light patterns directly within the configuration app in JavaScript. When adding a new light pattern, a boilerplate code pops up, which can be modified to program a light pattern. Using predefined colour variables within the code, these are subsequently automatically linked to the colour fields that can be adapted in the main view via the colour picker. Further, new \emph{scenario fields} can be added and suitable light patterns can be allocated to those, which will then be displayed in the main view. Finally, there is an option to hide the specific description of a scene into a generic enumeration (e.g. ``AV turning to the left'' to ``Scene 1''). This is to serve the purpose that the toolkit can be used in a setup where workshop participants have to guess the scenario based on the displayed light patterns.

\subsection{Implementation}
The \emph{simulation} component is implemented using the cross-platform game engine Unity\footnote{\url{https://unity.com/}, accessed August 2020}. Unity comes with a built-in feature to export projects across mobile platforms (e.g. iOS, Android), which allowed us to deploy the simulation app on the iPads. The different tablet views are all implemented within the same project and exported to a single app. This has been achieved through the Photon Unity Networking (PUN)\footnote{\url{https://www.photonengine.com/en-us/Photon}, accessed August 2020} package, which supports the development of multiplayer games. PUN synchronises all attributes of game objects (i.e. their position, rotation and scale) within a scene across the various players (e.g. in our case the different tablet views) in real time. For each tablet view, we only had to assign a new camera to a distinct game object. For the top views, to avoid any potential confusion for participants, we set the simulated AV invisible, as in our setup the model would be overlaid by the physical mock-up. Both the shared space and street scenario have been designed using a polygon styled city package downloaded from the Unity Asset Store\footnote{\url{https://assetstore.unity.com/packages/3d/environments/urban/city-package-107224}, accessed August 2020}.

We modelled the \emph{tangibles} in the 3D-modelling software Rhino\footnote{\url{https://www.rhino3d.com/}, accessed August 2020}, which were subsequently printed using white PLA material on an Ultimaker 3 Extended\footnote{\url{https://ultimaker.com/de/3d-printers/ultimaker-3}, accessed August 2020}. The AV model is composed of two parts which can be stacked together in order to house the electronic components and to retain access for exchanging the battery. We left slots in the AV model where we mounted three 7-pixel WS2812B RGB LED strips\footnote{\url{https://cdn-shop.adafruit.com/datasheets/WS2812B.pdf}, accessed August 2020}. The LEDs are connected to a WeMos D1 Mini board\footnote{\url{https://www.makerstore.com.au/product/elec-wemos-d1/}, accessed August 2020} which comes with an on-board wifi module and can be programmed using the Arduino development environment. The board and the LEDs were powered by a 9V block battery. To recognise the tangibles when placed on the tablet, we inserted three touch pen tips that we extracted from low-cost capacitive pens\footnote{\url{https://www.ebay.com.au/itm/10x-Mini-Capacitive-Stylus-Touch-Screen-Pen-For-Mobile-Phone-Tablet-iPad-iPhone-/201543592745}, accessed August 2020}. Aligning the touch pen pins spanning an irregular triangle and ensuring different spacing between the tips for the two tangibles facilitates identification of the objects and determining their position through a touch recognition script implemented within the Unity project.

The \emph{configuration app} was implemented as a native iOS application using the Cocoa Touch framework. For the animated light patterns, which can be developed within our app, we used the JavaScriptCore framework, which allows to evaluate JavaScript programs within iOS apps\footnote{\url{https://developer.apple.com/documentation/javascriptcore}, accessed August 2020}. Outsourcing the development and generation of the light patterns into the configuration app, has the advantage that there is no need to recompile the Unity application or reprogram the embedded hardware platform of the physical car prototype when new light patterns are being created. 

Additional to the network communication across the three simulation tablets, which comes out of the box through the Photon framework, our toolkit components require two additional network sockets: First, we used a TCP connection to communicate the position change of the physical car prototype to the configuration app. Second, to stream the light patterns in real-time from the configuration app to the low-res interface of the physical and simulated vehicle, we used DMX streams wrapped in an UDP package.

\section{Study Design}

We used the toolkit to inform the design of the low-res display attached to our real AV. At the same time, given the novel prototyping approach that we implemented, we were interested in understanding the value of using tangible objects and multiple displays. Both these objectives therefore became the driving forces for designing our study. As the emphasis of this paper is on the toolkit (rather than on the low-res display design), in this section we focus on those aspects of the study design that were relevant to the toolkit and its use during collaborative design exploration sessions. Below we describe the research questions, the study setup and materials, the participants and procedure, and the data analysis process. 

\subsection{Research Questions}
The following research questions informed the design of the study aspects reported in this paper: 
1) What is the value of combining tangible objects and multiple displays for evaluating AV-pedestrian interface design proposals with multiple stakeholders?
2) To what extent are participants able to quickly learn and use the toolkit? What are potential barriers?

\subsection{Setup and Materials}
\begin{figure}
  \includegraphics[width=1\linewidth]{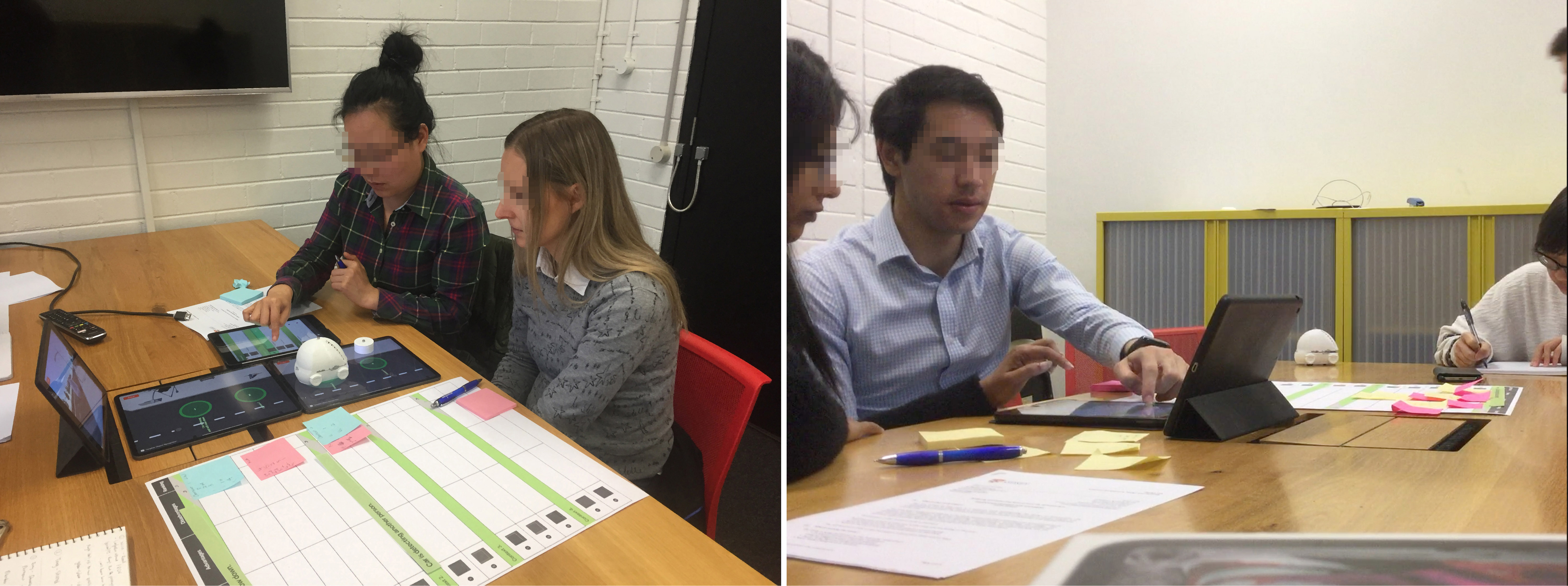}
  \caption{Participants interacting with the tangible multi-display representation (left), and the single-display simulation (right) in a collaborative design exploration session.}
  \Description{Participants interacting with the toolkit in the design exploration session.}
  \label{fig:user_study}
\end{figure}

The study was designed as a repeated measures experiment with the prototyping environment as independent variable: The study condition comprised the tangible multi-display toolkit, including the first-person view tablet, top-view tablet with tangibles, and the configuration app (see Fig. \ref{fig:user_study}, left). As a baseline condition, we used the first-person view tablet and the configuration app (right). We designed two different environments, a shared urban space and a street environment, and created four scenarios for each (see scenarios for shared environment in Fig. \ref{fig:toolkit_overview}). While the scenarios in each environment were similar, we slightly changed their order of appearance and designed different light patterns for the different environments. We used these two environments as we expected that people would have different expectations for how an AV would behave in a shared versus a road environment. Having two different environment further allowed us to compare the two conditions using a within-subject study design setup, where each participant experienced both conditions. 

Each session involved a pair of expert study participants. The tasks for the participants were to guess the scenarios based on the presented simulation and light patterns, to interpret the meaning of the light patterns, and to provide feedback on the design of the light patterns. Participants were encouraged to make changes to the colour schemes via the configuration app as part of their design exploration. At the end of each session, participants were asked to select their preferred set of light patterns across all four scenarios. To facilitate the process of providing feedback and generating design ideas that went beyond the restrictions of the implemented AV-interface, participants were given a design template (A2 portrait format) in each condition. The design template contained the light patterns presented in the configuration app and included several columns for participants to cluster sticky notes. All sessions were video and audio recorded with a GoPro camera.

\subsection{Participants and Procedure}
We conducted fourteen design exploration sessions in total (seven for each condition), with two participants per session. Therefore, we recruited fourteen participants (seven female) of various academic and professional backgrounds, covering different expertise that is considered relevant for the design of urban technologies \cite{Tomitsch2020, Malizia2018}. Participants' professions included two architects, one post-graduate architectural science student, one PhD researchers in architecture, two interaction designers, three PhD researchers in HCI, two PhD researchers in psychology, one software developer, one urbanist and one civil engineer working in transport planning. Each pair of participants was presented with both conditions and environments. Each pair was assigned one of the two conditions to start with; conditions and environments were pseudo-randomised to counterbalance for biasing effects of the prototyping representation and potential interaction effect of the presented environment.

\begin{figure}
  \includegraphics[width=1\linewidth]{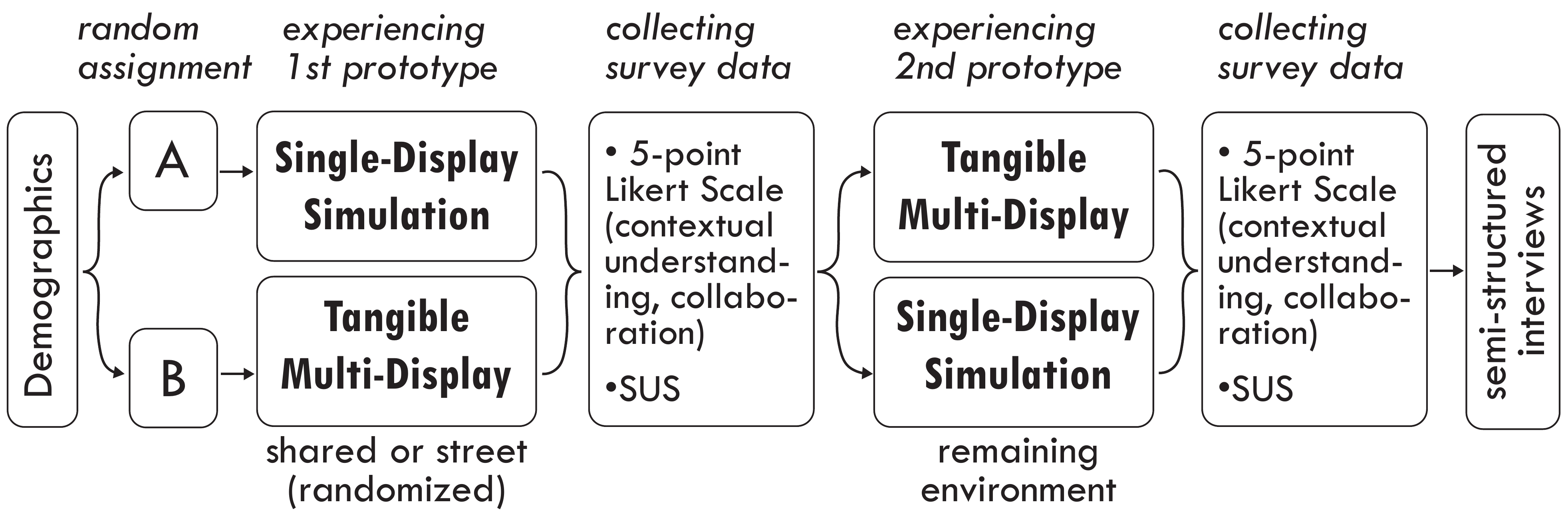}
  \caption{Study procedure and data collection.}
  \Description{Study procedure and data collection.}
  \label{fig:procedure}
\end{figure}

After giving informed consent, we started with a short introduction about shared autonomous mobility and AV-pedestrian interfaces. We then presented the main functionalities of the respective toolkit representation and introduced the tasks. After participants completed the tasks with their first assigned condition, we changed the toolkit condition and environment. After each condition, participants filled out a questionnaire on a five-point Likert scale with statements regarding collaboration and contextual understanding, and the System Usability Scale (SUS) \cite{brooke1996quick}. Each condition took approximately 35 minutes to complete. At the end of the second condition, we conducted a 10-minute interview with participants to discuss the toolkit. See Figure \ref{fig:procedure} for study procedure and data collection.

\subsection{Data Analysis}

For the purpose of this paper, we reviewed the transcripts from the exploratory design sessions, looking for data points (quotes) that related to one of our research questions. We then clustered the data points around the areas discussed in the following section. To get further insight on how our participants interacted with the toolkit and with each other, we reviewed the relevant video segments linked to the identified quotes. 

\section{Findings and Discussion}

In this section, we present the findings from the comparative evaluation of the toolkit and address the research questions outlined in the previous section. The section is structured to first discuss the findings related to the collaborative design exploration process, followed by a reflection on the usability of the toolkit and a discussion of the toolkit design. 

\subsection{Collaborative Design Exploration process}

\begin{figure}
  \includegraphics[width=0.95\linewidth]{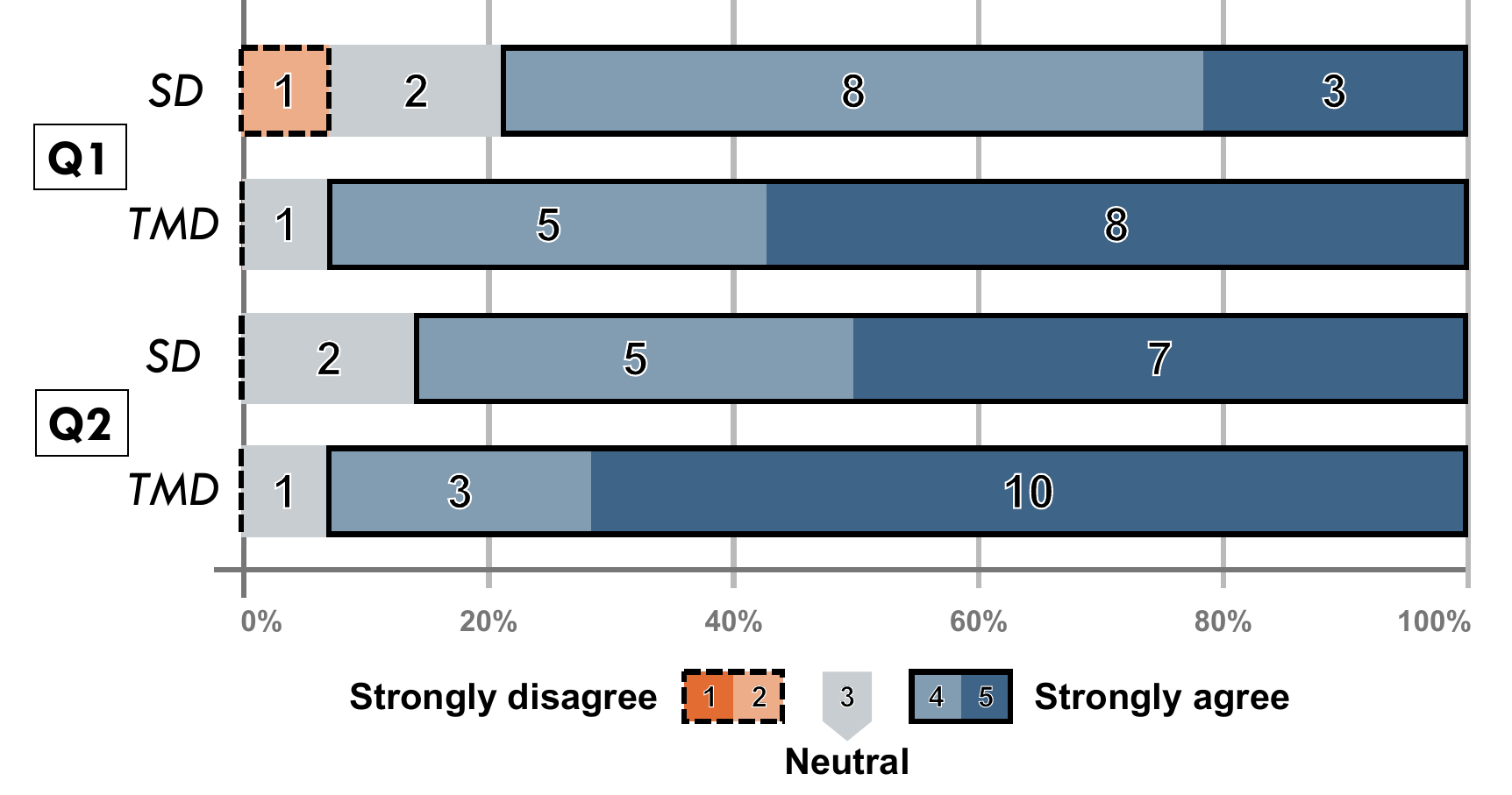}
  \caption{Results of Q1: ``The prototyping toolkit facilitated the understanding of the design context'' and Q2: ``The prototyping toolkit facilitated communication''. \textit{[SD=Single Display Simulation, TMD=Tangible Multi-Display Toolkit]}}
  \Description{Results regarding understanding of design context and facilitation of communication.}
  \label{fig:likert}
\end{figure}

\subsubsection{Supporting participants' understanding of the context}
Previous research has highlighted the importance of context-based methods for prototyping and evaluating interactions with AVs \cite{Flohr2020}, in particular since people still are not yet able to build on personal experiences with these kinds of technologies \cite{Philipsen2019}. While both representations received a positive response to support understanding the context for which the light patterns were designed for, there was a preference towards the tangible multi-display representation (median=5.0, mode=5.0, see Fig. \ref{fig:likert}: Q1) over the single display representation (median=4.0, mode=4.0).\footnote{We report the median and mode values here as they are a useful indicator for descriptive quantitative analysis that aims to indicate a tendency rather than significant effects.} One participant (P3) stated that interacting with the tangible multi-display representation and being able to change the position \textit{``better reflected a real-life experience as a pedestrian, in which [she] would also move around''}. Seeing the top-view perspective and ego-perspective simultaneously \textit{``gave a more holistic picture''} (P9) of the situation. Another participant (P10) referred to the top-view perspective and mentioned that it was helpful \textit{``to see spatially where the scenarios took place on the map''}. Further, being able to \textit{``change positions and perspectives''} helped to quickly re-asses the AV-pedestrian interface from various pedestrian point of views (P10), which was considered as \textit{``helpful to understand limitations of the design''} (P5), for example, in terms of the positioning of the lights on the vehicle. We also observed that participants would often jump between various scenarios, in order to see if their current selection of a light pattern would be consistent with patterns that they would have selected for previous scenarios. Having the ability to directly control the simulation through the tangible objects facilitated this rapid exploration and comparison by moving back and forth between different scenarios. In terms of a more holistic consolidation of contextual aspects, participants also stated that for quickly evaluating the efficacy of the AV-pedestrian interface in various situations, it would have been helpful being able to simulate various traffic conditions (e.g. low vs. high) and lighting conditions (e.g. daylight vs. low light).

\subsubsection{Facilitating communication between participants} 
Besides evaluation and testing, an important purpose of prototypes is to facilitate communication between various stakeholders \cite{Lim2008}. Both of the prototype representations used in the two toolkit conditions received a positive response in terms of the facilitation of communication, with a slightly higher rating for the multi-display toolkit (median=5.0, mode=5.0, see Fig. \ref{fig:likert}: Q2) over the single-display simulation (median=4.5, mode=5.0). In retrospect, participants stated that the tangible multi-display toolkit was more engaging, with the embodied and interactive representation \textit{``facilitat[ing] conversations''} and \textit{``making it easier to describe things''} (P3). The higher level of engagement was also supported through the multiple interactive means of the tangible multi-display representation: While in the single-display setup, mostly one person was taking responsibility to configure the light patterns and select the scenarios, in the multi-display setup we observed collective interactions on the top-view tablets. Besides the toolkit itself the printed design template was also considered as important to keep up the communication flow and to have a workspace to share ideas through adding sticky notes: \textit{``Everyone was able to see the [design template] whereas only one of us could see what was on the iPad [configuration app]''} (P6).

\begin{figure}
  \includegraphics[width=0.95\linewidth]{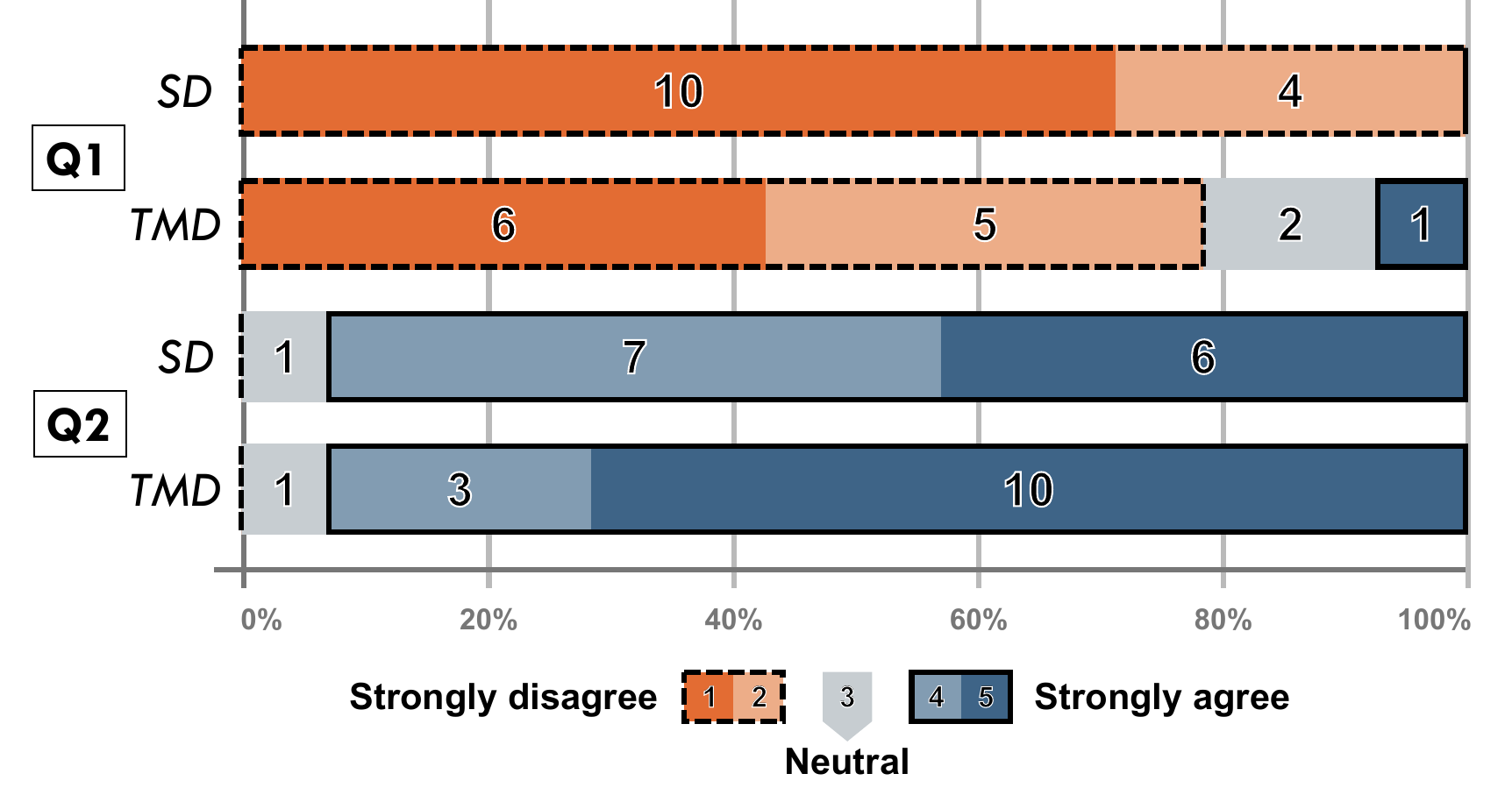}
  \caption{Results of Q1: ``I needed to learn a lot of things before going with the toolkit'' and Q2: ``I would imagine that most people would learn to use the toolkit very quickly''. \textit{[SD=Single Display Simulation, TMD =Tangible Multi-Display Toolkit]}}
  \Description{Results from the SUS scale.}
  \label{fig:sus}
\end{figure}

\subsection{Usability of the Toolkit}
To successfully involve stakeholders of various backgrounds in the design process of cyber-physical systems, it is important that the methods and tools for engaging people, who are not familiar with prototyping techniques, are easy to understand and use. Hence, we evaluated the usability of the toolkit itself as this could have a potential effect on participants' ability to successfully use the toolkit as part of a collaborative design exploration process. To investigate the usability of the toolkit, we included two questions from the SUS questionnaire \cite{brooke1996quick} in the post-study questionnaire (Fig. \ref{fig:sus}). We further looked for post-study interview quotes that included references to the usability of the toolkit itself -- both for the baseline and the study condition.  

Interview quotes and responses from the questionnaire indicated that after an initial phase of exploring the setups used in both conditions, participants felt comfortable interacting with the tools used in each of the conditions. While the tangible multi-display toolkit required slightly more investment to explore all its components, participants' responses also indicated a faster learning curve here (see Fig. \ref{fig:sus}). Participants stated that \textit{``it took a little bit longer to set up [the tangible multi-display toolkit] and understand [the components]''} (P6). However, after \textit{``feeling a bit overwhelmed at the beginning''} it was considered as \textit{``more engaging''} (P8). Participants also stated that they \textit{``liked the [ability] of adjusting the colours''} (P13), but that they would have wished more tools to make adjustments to the AV-pedestrian interface, for example, allowing for \textit{``changing the [animation patterns] of the lights''} (P5).


\subsection{Lessons Learnt: Toolkit Design}

Below we reflect on lessons learnt in terms of the design of the tangible multi-display toolkit. We hope that these lessons are able to offer guidance for others adopting a similar approach for engaging multiple stakeholders in early collaborative design exploration of AV-pedestrian interfaces and other cyber-physical systems. 

\subsubsection{Multi-Perspective Prototyping} 
We found that providing participants multiple perspectives of the prototype representation helped them obtain a more holistic understanding of the context. The tangible objects for interacting with the simulated environment thereby reinforced the connection between the two simulated perspectives and even simulated a sense of immersion. One participant (P5) drew similarities to the sense of \textit{``embodiment in virtual reality environments''}, stating that \textit{``[she] felt like standing in the scene}, which was reinforced \textit{``[through] the scene moving with [her]''} by interacting with the tangibles. Simulating a sense of immersion through multi-perspective prototyping provides benefits in the context of collaborative design explorations: thus, compared to immersive VR simulations, multiple participants can experience the prototype simultaneously, talking through design concepts without feeling disconnect from each other \cite{Radianti2020}, and incorporate physical materials (e.g. post-it notes) in ideation tasks \cite{Jensen2018}.

We were able to achieve this sense of immersion and holistic understanding of the context through relative simple means embodied in our toolkit: 1) Displaying two perspectives of the simulated environment on multiple displays; 2) using separate screens, instead of for example showing a split-view representation on one screen, enabling us to prototype a pseudo three-dimensional setup by orienting the displays in different directions (horizontal and vertical); and 3) using the Photon framework in Unity, which offers out-of-the box synchronisation across multiple instances of the application $-$ this could also be useful for a wide range of other prototyping contexts, such as simulating in-vehicle experiences and those outside of the car simultaneously.



\subsubsection{Mixed Material Prototyping} 
Using a mixed material prototyping environment, has demonstrated that it can provide value to cover various aspects of the envisioned AV-pedestrian interface, including its aesthetic appearance and interplay with the larger system and environment. Our participants stated that \textit{``the light design was clearer to perceive on the model''} (P6), rather than \textit{``just seeing it on the iPad''} (P10). On the other hand, they mentioned the importance to \textit{``assess the light signals together with the movement of the vehicle and the surrounding environment''} (P10), which was represented in the simulation. Revisiting video recordings of the design sessions confirmed that participants used both the low-res interface on the physical model and the 3D-modelled low-res display in the simulation to assess the light patterns.

To enable efficient mixed reality prototyping, we recommend our approach of outsourcing the main interface components under investigation. Generating the light patterns on the configuration app enabled us to quickly develop new light patterns without making any changes to the Unity application and the code running on the Arduino board. However, separating interface elements from simulated system and environment, can also lead to issues in terms of synchronisation. For example, participants stated that for some scenarios, the dynamic light patterns and simulation of the system (e.g. motion of the vehicle) were off set, which needs to be improved through implementing additional synchronisation routines. Further, participants stated in regards to the looping of the dynamic light patterns, it would be helpful to add a delay at the beginning to better reinforce the sequential order of the animations.

\subsubsection{Augmented Scale Scenario Prototyping}
Previous work on exploratory interaction design techniques for human vehicle interaction has reported on the successful employment of scale scenario prototyping in their design process \cite{Pettersson2017}. 
In our approach, we augmented scale scenario prototyping through additional digital elements, such as the light patterns and simulated scenes. While still retaining the tangible character of scale scenarios, our approach allows for dynamic aspects of the interface itself (e.g. moving light patterns) to be simulated, while also catering for the dynamic and complex situations that can occur in urban environments, in which robots and AVs will operate in (e.g. pedestrian walking in front of the vehicle). While the current version of our toolkit is comprised of only two tangible objects, it would be possible to extend the palette to allow for a more interactive simulation of dynamic situations, such as simulating interactions between an AV and multiple pedestrians or bicyclists.

Although the scale scenario approach within our prototyping environment has contributed towards sense-making of the design context at large, there are also potential pitfalls that need to be considered: for example, two of our workshop participants in the beginning considered the small-scale car prototype as a self-contained and integral artefact itself. Based on their mental model of how such artefacts should behave, they assumed that the light patterns on the vehicle would indicate a low-battery level of the artefact itself rather than drawing a connection to AV-interfaces. This hints towards the need for careful consideration of how to abstract envisioned systems through prototypes, and also shows that prototypical representations convey additional underlying meaning.



\section{Conclusion}

The advent of cyber-physical systems and their expected deployment in real-world urban environments, brings a range of new challenges into the domain of interaction design, such as, how to design interactions between people and AVs as an example of a technology that will likely become available in the near future yet designers and prospective users are not able to draw on their own experiences of using these technologies. To support the early exploration of such systems and their user interfaces, we presented a novel approach using a \textit{tangible multi-display toolkit}. Our approach lowers the barrier to involve a wide range of stakeholders and to engage them collaboratively in the evaluation of interface design proposals. In response to the observation that HCI toolkits face challenges with their adoption by others \cite{Ledo2018}, we used the paper to foreground the value of our toolkit as an approach. In other words, in addition to describing how we implemented the toolkit, we reported on the design and usage of the various toolkit components, thus enabling replication and adoption by others. We therefore hope that our approach is more broadly applicable to prototype and evaluate context-based interfaces, such as for urban robots or smart environments.

The value of the toolkit as an approach is illustrated through our findings from a comparative evaluation of the toolkit in design sessions with experts. During these sessions, pairs of experts evaluated different interface design solutions. Data collected during the sessions and from subsequent questionnaires and interviews, indicate that our approach supported participants in developing a more holistic understanding of the design context and facilitated communication between participants. Based on our findings, we suggest that a combination of tangible objects and multiple displays offers effective means for evaluating early interface design concepts. As such, the toolkit approach outlined in this paper fills a gap as it is able to create an immersive experience by simulating multiple perspectives and allowing participants to directly control the displayed perspectives and scenarios. This meets similar objectives that have been attributed to using virtual reality setups for evaluating AV-pedestrian interfaces, while at the same time allowing multiple participants to collaboratively evaluate and discuss the concepts. 

Our findings also highlight avenues for future work, such as the development of additional support tools that allow participants to co-create potential solutions. Our study was limited in this regard, as our aim was to evaluate a series of interface design proposals. While the toolkit allowed participants to adapt certain aspects, such as the colour used in the low-res lighting display, our aim was not to enable a truly participatory design process. 

The paper contributes to the growing body of work across HCI, human-machine interaction and interaction design that focuses on approaches for prototyping and evaluating cyber-physical systems. These approaches are critical to advance these types of systems not just from a technical perspective but, importantly, from a human-centred perspective. Beyond considering the end user, these systems rely heavily on the input from multiple stakeholders with expertise in different areas to cover not only the complexity of the interfaces themselves but also the complexity of the urban environment in which they are being deployed. In the words of one of our participants, toolkits like the one presented in this paper enable designers and expert stakeholders to collaboratively \textit{``imagine this world''} that has yet to be built. 


\begin{acks}
The authors acknowledge the statistical assistance of Kathrin Schemann of the Sydney Informatics Hub, a Core Research Facility of the University of Sydney. This research was supported partially by the Sydney Institute for Robotics and Intelligent Systems (SIRIS) and ARC Discovery Project DP200102604 Trust and Safety in Autonomous Mobility Systems: A Human-centred Approach.
\end{acks}

\balance{}
\balance{}

\bibliographystyle{ACM-Reference-Format}
\bibliography{sample-base}


\begin{thebibliography}{48}


\ifx \showCODEN    \undefined \def \showCODEN     #1{\unskip}     \fi
\ifx \showDOI      \undefined \def \showDOI       #1{#1}\fi
\ifx \showISBNx    \undefined \def \showISBNx     #1{\unskip}     \fi
\ifx \showISBNxiii \undefined \def \showISBNxiii  #1{\unskip}     \fi
\ifx \showISSN     \undefined \def \showISSN      #1{\unskip}     \fi
\ifx \showLCCN     \undefined \def \showLCCN      #1{\unskip}     \fi
\ifx \shownote     \undefined \def \shownote      #1{#1}          \fi
\ifx \showarticletitle \undefined \def \showarticletitle #1{#1}   \fi
\ifx \showURL      \undefined \def \showURL       {\relax}        \fi
\providecommand\bibfield[2]{#2}
\providecommand\bibinfo[2]{#2}
\providecommand\natexlab[1]{#1}
\providecommand\showeprint[2][]{arXiv:#2}

\bibitem[\protect\citeauthoryear{Benderius, Berger, and Lundgren}{Benderius
  et~al\mbox{.}}{2018}]%
        {Benderius2018}
\bibfield{author}{\bibinfo{person}{Ola Benderius}, \bibinfo{person}{Christian
  Berger}, {and} \bibinfo{person}{Victor~Malmsten Lundgren}.}
  \bibinfo{year}{2018}\natexlab{}.
\newblock \showarticletitle{The Best Rated Human–Machine Interface Design for
  Autonomous Vehicles in the 2016 Grand Cooperative Driving Challenge}.
\newblock \bibinfo{journal}{\emph{IEEE Transactions on Intelligent
  Transportation Systems}} \bibinfo{volume}{19}, \bibinfo{number}{4}
  (\bibinfo{year}{2018}), \bibinfo{pages}{1302--1307}.
\newblock


\bibitem[\protect\citeauthoryear{Boll, Koelle, and Cauchard}{Boll
  et~al\mbox{.}}{2019}]%
        {Boll2019}
\bibfield{author}{\bibinfo{person}{Susanne Boll}, \bibinfo{person}{Marion
  Koelle}, {and} \bibinfo{person}{Jessica Cauchard}.}
  \bibinfo{year}{2019}\natexlab{}.
\newblock \showarticletitle{{Understanding the Socio-Technical Impact of
  Automated (Aerial) Vehicles on Casual Bystanders}}. In
  \bibinfo{booktitle}{\emph{{1st International Workshop on Human-Drone
  Interaction}}}. {Ecole Nationale de l'Aviation Civile [ENAC]},
  \bibinfo{address}{Glasgow, United Kingdom}.
\newblock
\urldef\tempurl%
\url{https://hal.archives-ouvertes.fr/hal-02128379}
\showURL{%
\tempurl}


\bibitem[\protect\citeauthoryear{Brooke}{Brooke}{1996}]%
        {brooke1996quick}
\bibfield{author}{\bibinfo{person}{John Brooke}.}
  \bibinfo{year}{1996}\natexlab{}.
\newblock \bibinfo{booktitle}{\emph{"SUS-A quick and dirty usability scale."
  Usability evaluation in industry}}.
\newblock \bibinfo{publisher}{CRC Press}.
\newblock
\urldef\tempurl%
\url{https://www.crcpress.com/product/isbn/9780748404605}
\showURL{%
\tempurl}
\newblock
\shownote{ISBN: 9780748404605.}


\bibitem[\protect\citeauthoryear{Broy, Lindner, and Alt}{Broy
  et~al\mbox{.}}{2016}]%
        {Broy2016}
\bibfield{author}{\bibinfo{person}{Nora Broy}, \bibinfo{person}{Verena
  Lindner}, {and} \bibinfo{person}{Florian Alt}.}
  \bibinfo{year}{2016}\natexlab{}.
\newblock \showarticletitle{The S3D-UI Designer: Creating User Interface
  Prototypes for 3D Displays}. In \bibinfo{booktitle}{\emph{Proceedings of the
  15th International Conference on Mobile and Ubiquitous Multimedia}}
  (Rovaniemi, Finland) \emph{(\bibinfo{series}{MUM ’16})}.
  \bibinfo{publisher}{Association for Computing Machinery},
  \bibinfo{address}{New York, NY, USA}, \bibinfo{pages}{49–55}.
\newblock
\showISBNx{9781450348607}
\urldef\tempurl%
\url{https://doi.org/10.1145/3012709.3012727}
\showDOI{\tempurl}


\bibitem[\protect\citeauthoryear{Broy, Schneegass, Alt, and Schmidt}{Broy
  et~al\mbox{.}}{2014}]%
        {Broy2014}
\bibfield{author}{\bibinfo{person}{Nora Broy}, \bibinfo{person}{Stefan
  Schneegass}, \bibinfo{person}{Florian Alt}, {and} \bibinfo{person}{Albrecht
  Schmidt}.} \bibinfo{year}{2014}\natexlab{}.
\newblock \showarticletitle{FrameBox and MirrorBox: Tools and Guidelines to
  Support Designers in Prototyping Interfaces for 3D Displays}. In
  \bibinfo{booktitle}{\emph{Proceedings of the SIGCHI Conference on Human
  Factors in Computing Systems}} (Toronto, Ontario, Canada)
  \emph{(\bibinfo{series}{CHI ’14})}. \bibinfo{publisher}{Association for
  Computing Machinery}, \bibinfo{address}{New York, NY, USA},
  \bibinfo{pages}{2037–2046}.
\newblock
\showISBNx{9781450324731}
\urldef\tempurl%
\url{https://doi.org/10.1145/2556288.2557183}
\showDOI{\tempurl}


\bibitem[\protect\citeauthoryear{Buxton}{Buxton}{2007}]%
        {Buxton2007}
\bibfield{author}{\bibinfo{person}{Bill Buxton}.}
  \bibinfo{year}{2007}\natexlab{}.
\newblock \bibinfo{booktitle}{\emph{Sketching User Experiences: Getting the
  Design Right and the Right Design}}.
\newblock \bibinfo{publisher}{Morgan Kaufmann Publishers Inc.},
  \bibinfo{address}{San Francisco, CA, USA}.
\newblock
\showISBNx{0123740371}


\bibitem[\protect\citeauthoryear{Cila, Smit, Giaccardi, and Kr\"{o}se}{Cila
  et~al\mbox{.}}{2017}]%
        {Nazli2017}
\bibfield{author}{\bibinfo{person}{Nazli Cila}, \bibinfo{person}{Iskander
  Smit}, \bibinfo{person}{Elisa Giaccardi}, {and} \bibinfo{person}{Ben
  Kr\"{o}se}.} \bibinfo{year}{2017}\natexlab{}.
\newblock \showarticletitle{Products as Agents: Metaphors for Designing the
  Products of the IoT Age}. In \bibinfo{booktitle}{\emph{Proceedings of the
  2017 CHI Conference on Human Factors in Computing Systems}} (Denver,
  Colorado, USA) \emph{(\bibinfo{series}{CHI ’17})}.
  \bibinfo{publisher}{Association for Computing Machinery},
  \bibinfo{address}{New York, NY, USA}, \bibinfo{pages}{448–459}.
\newblock
\showISBNx{9781450346559}
\urldef\tempurl%
\url{https://doi.org/10.1145/3025453.3025797}
\showDOI{\tempurl}


\bibitem[\protect\citeauthoryear{de~Clercq, Dietrich, Velasco, de~Winter, and
  Happee}{de~Clercq et~al\mbox{.}}{2019}]%
        {Clercq2019}
\bibfield{author}{\bibinfo{person}{Koen de Clercq}, \bibinfo{person}{Andre
  Dietrich}, \bibinfo{person}{Juan Pablo~Núñez Velasco},
  \bibinfo{person}{Joost de Winter}, {and} \bibinfo{person}{Riender Happee}.}
  \bibinfo{year}{2019}\natexlab{}.
\newblock \showarticletitle{External Human-Machine Interfaces on Automated
  Vehicles: Effects on Pedestrian Crossing Decisions}.
\newblock \bibinfo{journal}{\emph{Human Factors}} \bibinfo{volume}{61},
  \bibinfo{number}{8} (\bibinfo{year}{2019}), \bibinfo{pages}{1353--1370}.
\newblock
\urldef\tempurl%
\url{https://doi.org/10.1177/0018720819836343}
\showDOI{\tempurl}
\showeprint{https://doi.org/10.1177/0018720819836343}
\newblock
\shownote{PMID: 30912985.}


\bibitem[\protect\citeauthoryear{Dey, Habibovic, Löcken, Wintersberger,
  Pfleging, Riener, Martens, and Terken}{Dey et~al\mbox{.}}{2020a}]%
        {Dey2020}
\bibfield{author}{\bibinfo{person}{Debargha Dey}, \bibinfo{person}{Azra
  Habibovic}, \bibinfo{person}{Andreas Löcken}, \bibinfo{person}{Philipp
  Wintersberger}, \bibinfo{person}{Bastian Pfleging}, \bibinfo{person}{Andreas
  Riener}, \bibinfo{person}{Marieke Martens}, {and} \bibinfo{person}{Jacques
  Terken}.} \bibinfo{year}{2020}\natexlab{a}.
\newblock \showarticletitle{Taming the eHMI jungle: A classification taxonomy
  to guide, compare, and assess the design principles of automated vehicles'
  external human-machine interfaces}.
\newblock \bibinfo{journal}{\emph{Transportation Research Interdisciplinary
  Perspectives}}  \bibinfo{volume}{7} (\bibinfo{year}{2020}),
  \bibinfo{pages}{100174}.
\newblock
\showISSN{2590-1982}
\urldef\tempurl%
\url{https://doi.org/10.1016/j.trip.2020.100174}
\showDOI{\tempurl}


\bibitem[\protect\citeauthoryear{Dey, Habibovic, Pfleging, Martens, and
  Terken}{Dey et~al\mbox{.}}{2020b}]%
        {Debargha2020}
\bibfield{author}{\bibinfo{person}{Debargha Dey}, \bibinfo{person}{Azra
  Habibovic}, \bibinfo{person}{Bastian Pfleging}, \bibinfo{person}{Marieke
  Martens}, {and} \bibinfo{person}{Jacques Terken}.}
  \bibinfo{year}{2020}\natexlab{b}.
\newblock \showarticletitle{Color and Animation Preferences for a Light Band
  EHMI in Interactions Between Automated Vehicles and Pedestrians}. In
  \bibinfo{booktitle}{\emph{Proceedings of the 2020 CHI Conference on Human
  Factors in Computing Systems}} (Honolulu, HI, USA)
  \emph{(\bibinfo{series}{CHI ’20})}. \bibinfo{publisher}{Association for
  Computing Machinery}, \bibinfo{address}{New York, NY, USA},
  \bibinfo{pages}{1–13}.
\newblock
\showISBNx{9781450367080}
\urldef\tempurl%
\url{https://doi.org/10.1145/3313831.3376325}
\showDOI{\tempurl}


\bibitem[\protect\citeauthoryear{Eisma, van Bergen, ter Brake, Hensen,
  Tempelaar, and de~Winter}{Eisma et~al\mbox{.}}{2020}]%
        {Eisma2020}
\bibfield{author}{\bibinfo{person}{Y.B. Eisma}, \bibinfo{person}{Steven van
  Bergen}, \bibinfo{person}{Sjoerd ter Brake}, \bibinfo{person}{Matthijs
  Hensen}, \bibinfo{person}{Willem~Jan Tempelaar}, {and}
  \bibinfo{person}{J.C.F. de Winter}.} \bibinfo{year}{2020}\natexlab{}.
\newblock \showarticletitle{External Human–Machine Interfaces: The Effect of
  Display Location on Crossing Intentions and Eye Movements}.
\newblock \bibinfo{journal}{\emph{Information}} \bibinfo{volume}{11},
  \bibinfo{number}{1} (\bibinfo{year}{2020}).
\newblock
\showISSN{2078-2489}
\urldef\tempurl%
\url{https://doi.org/10.3390/info11010013}
\showDOI{\tempurl}


\bibitem[\protect\citeauthoryear{Flohr, Janetzko, Wallach, Scholz, and
  Kr\"{u}ger}{Flohr et~al\mbox{.}}{2020}]%
        {Flohr2020}
\bibfield{author}{\bibinfo{person}{Lukas~A. Flohr}, \bibinfo{person}{Dominik
  Janetzko}, \bibinfo{person}{Dieter~P. Wallach}, \bibinfo{person}{Sebastian~C.
  Scholz}, {and} \bibinfo{person}{Antonio Kr\"{u}ger}.}
  \bibinfo{year}{2020}\natexlab{}.
\newblock \showarticletitle{Context-Based Interface Prototyping and Evaluation
  for (Shared) Autonomous Vehicles Using a Lightweight Immersive Video-Based
  Simulator}. In \bibinfo{booktitle}{\emph{Proceedings of the 2020 ACM
  Designing Interactive Systems Conference}} (Eindhoven, Netherlands)
  \emph{(\bibinfo{series}{DIS ’20})}. \bibinfo{publisher}{Association for
  Computing Machinery}, \bibinfo{address}{New York, NY, USA},
  \bibinfo{pages}{1379–1390}.
\newblock
\showISBNx{9781450369749}
\urldef\tempurl%
\url{https://doi.org/10.1145/3357236.3395468}
\showDOI{\tempurl}


\bibitem[\protect\citeauthoryear{Forshaw, Cruickshank, and Dix}{Forshaw
  et~al\mbox{.}}{2012}]%
        {Forshaw2012}
\bibfield{author}{\bibinfo{person}{Stephen Forshaw}, \bibinfo{person}{Leon
  Cruickshank}, {and} \bibinfo{person}{Alan Dix}.}
  \bibinfo{year}{2012}\natexlab{}.
\newblock \showarticletitle{Collaborative Communication Tools for Designing:
  Physical-Cyber Environments}.
\newblock
\urldef\tempurl%
\url{https://doi.org/10.14236/ewic/HCI2012.93}
\showDOI{\tempurl}


\bibitem[\protect\citeauthoryear{Garro, Vaccaro, Dutr\'{e}, and Stegen}{Garro
  et~al\mbox{.}}{2019}]%
        {Garro2019}
\bibfield{author}{\bibinfo{person}{Alfredo Garro}, \bibinfo{person}{Vittorio
  Vaccaro}, \bibinfo{person}{Stefan Dutr\'{e}}, {and} \bibinfo{person}{Jef
  Stegen}.} \bibinfo{year}{2019}\natexlab{}.
\newblock \showarticletitle{Cyber-Physical Systems Engineering: Model-Based
  Solutions}. In \bibinfo{booktitle}{\emph{Proceedings of the 2019 Summer
  Simulation Conference}} (Berlin, Germany) \emph{(\bibinfo{series}{SummerSim
  ’19})}. \bibinfo{publisher}{Society for Computer Simulation International},
  \bibinfo{address}{San Diego, CA, USA}, Article \bibinfo{articleno}{46},
  \bibinfo{numpages}{12}~pages.
\newblock


\bibitem[\protect\citeauthoryear{Gerber, Schroeter, and Vehns}{Gerber
  et~al\mbox{.}}{2019}]%
        {Gerber2019}
\bibfield{author}{\bibinfo{person}{Michael~A. Gerber}, \bibinfo{person}{Ronald
  Schroeter}, {and} \bibinfo{person}{Julia Vehns}.}
  \bibinfo{year}{2019}\natexlab{}.
\newblock \showarticletitle{A Video-Based Automated Driving Simulator for
  Automotive UI Prototyping, UX and Behaviour Research}. In
  \bibinfo{booktitle}{\emph{Proceedings of the 11th International Conference on
  Automotive User Interfaces and Interactive Vehicular Applications}} (Utrecht,
  Netherlands) \emph{(\bibinfo{series}{AutomotiveUI '19})}.
  \bibinfo{publisher}{Association for Computing Machinery},
  \bibinfo{address}{New York, NY, USA}, \bibinfo{pages}{14–23}.
\newblock
\showISBNx{9781450368841}
\urldef\tempurl%
\url{https://doi.org/10.1145/3342197.3344533}
\showDOI{\tempurl}


\bibitem[\protect\citeauthoryear{Harrison, Horstman, Hsieh, and
  Hudson}{Harrison et~al\mbox{.}}{2012}]%
        {Harrison2012}
\bibfield{author}{\bibinfo{person}{Chris Harrison}, \bibinfo{person}{John
  Horstman}, \bibinfo{person}{Gary Hsieh}, {and} \bibinfo{person}{Scott
  Hudson}.} \bibinfo{year}{2012}\natexlab{}.
\newblock \showarticletitle{Unlocking the Expressivity of Point Lights}. In
  \bibinfo{booktitle}{\emph{Proceedings of the SIGCHI Conference on Human
  Factors in Computing Systems}} (Austin, Texas, USA)
  \emph{(\bibinfo{series}{CHI ’12})}. \bibinfo{publisher}{Association for
  Computing Machinery}, \bibinfo{address}{New York, NY, USA},
  \bibinfo{pages}{1683–1692}.
\newblock
\showISBNx{9781450310154}
\urldef\tempurl%
\url{https://doi.org/10.1145/2207676.2208296}
\showDOI{\tempurl}


\bibitem[\protect\citeauthoryear{Henderson, Tomitsch, and Leong}{Henderson
  et~al\mbox{.}}{2018}]%
        {Henderson2018}
\bibfield{author}{\bibinfo{person}{Hamish Henderson}, \bibinfo{person}{Martin
  Tomitsch}, {and} \bibinfo{person}{Tuck~Wah Leong}.}
  \bibinfo{year}{2018}\natexlab{}.
\newblock \showarticletitle{Tools to Think with: Augmenting User Interviews
  with Rapid Modular Prototypes}. In \bibinfo{booktitle}{\emph{Proceedings of
  the 30th Australian Conference on Computer-Human Interaction}} (Melbourne,
  Australia) \emph{(\bibinfo{series}{OzCHI ’18})}.
  \bibinfo{publisher}{Association for Computing Machinery},
  \bibinfo{address}{New York, NY, USA}, \bibinfo{pages}{256–260}.
\newblock
\showISBNx{9781450361880}
\urldef\tempurl%
\url{https://doi.org/10.1145/3292147.3292209}
\showDOI{\tempurl}


\bibitem[\protect\citeauthoryear{Hoggenmueller and Wiethoff}{Hoggenmueller and
  Wiethoff}{2015}]%
        {Hoggenmueller2015}
\bibfield{author}{\bibinfo{person}{Marius Hoggenmueller} {and}
  \bibinfo{person}{Alexander Wiethoff}.} \bibinfo{year}{2015}\natexlab{}.
\newblock \showarticletitle{Blinking Lights and Other Revelations: Experiences
  Designing Hybrid Media Fa\c{c}Ades}. In \bibinfo{booktitle}{\emph{Proceedings
  of the 4th International Symposium on Pervasive Displays}} (Saarbruecken,
  Germany) \emph{(\bibinfo{series}{PerDis ’15})}.
  \bibinfo{publisher}{Association for Computing Machinery},
  \bibinfo{address}{New York, NY, USA}, \bibinfo{pages}{77–82}.
\newblock
\showISBNx{9781450336086}
\urldef\tempurl%
\url{https://doi.org/10.1145/2757710.2757725}
\showDOI{\tempurl}


\bibitem[\protect\citeauthoryear{Holl\"{a}nder, Colley, Mai, H\"{a}kkil\"{a},
  Alt, and Pfleging}{Holl\"{a}nder et~al\mbox{.}}{2019}]%
        {Hollaender2019}
\bibfield{author}{\bibinfo{person}{Kai Holl\"{a}nder}, \bibinfo{person}{Ashley
  Colley}, \bibinfo{person}{Christian Mai}, \bibinfo{person}{Jonna
  H\"{a}kkil\"{a}}, \bibinfo{person}{Florian Alt}, {and}
  \bibinfo{person}{Bastian Pfleging}.} \bibinfo{year}{2019}\natexlab{}.
\newblock \showarticletitle{Investigating the Influence of External Car
  Displays on Pedestrians' Crossing Behavior in Virtual Reality}. In
  \bibinfo{booktitle}{\emph{Proceedings of the 21st International Conference on
  Human-Computer Interaction with Mobile Devices and Services}} (Taipei,
  Taiwan) \emph{(\bibinfo{series}{MobileHCI '19})}.
  \bibinfo{publisher}{Association for Computing Machinery},
  \bibinfo{address}{New York, NY, USA}, Article \bibinfo{articleno}{27},
  \bibinfo{numpages}{11}~pages.
\newblock
\showISBNx{9781450368254}
\urldef\tempurl%
\url{https://doi.org/10.1145/3338286.3340138}
\showDOI{\tempurl}


\bibitem[\protect\citeauthoryear{Jensen, Thiel, Hoggan, and Bødker}{Jensen
  et~al\mbox{.}}{2018}]%
        {Jensen2018}
\bibfield{author}{\bibinfo{person}{Mads~Møller Jensen},
  \bibinfo{person}{Sarah-Kristin Thiel}, \bibinfo{person}{Eve Hoggan}, {and}
  \bibinfo{person}{Susanne Bødker}.} \bibinfo{year}{2018}\natexlab{}.
\newblock \showarticletitle{Physical Versus Digital Sticky Notes in
  Collaborative Ideation}. In \bibinfo{booktitle}{\emph{Computer Supported
  Cooperative Work (CSCW)}}.
\newblock


\bibitem[\protect\citeauthoryear{Kray, Olivier, Guo, Singh, Ha, and
  Blythe}{Kray et~al\mbox{.}}{2007}]%
        {Kray2007}
\bibfield{author}{\bibinfo{person}{Christian Kray}, \bibinfo{person}{Patrick
  Olivier}, \bibinfo{person}{Amy~Weihong Guo}, \bibinfo{person}{Pushpendra
  Singh}, \bibinfo{person}{Hai~Nam Ha}, {and} \bibinfo{person}{Phil Blythe}.}
  \bibinfo{year}{2007}\natexlab{}.
\newblock \showarticletitle{Taming context: A key challenge in evaluating the
  usability of ubiquitous systems}. In \bibinfo{booktitle}{\emph{Ubiquitous
  Systems Evaluation 2007 (USE '07) - Workshop at Ubicomp 2007}}.
\newblock


\bibitem[\protect\citeauthoryear{Kun, Boll, and Schmidt}{Kun
  et~al\mbox{.}}{2016}]%
        {Kun2016}
\bibfield{author}{\bibinfo{person}{Andrew~L. Kun}, \bibinfo{person}{Susanne
  Boll}, {and} \bibinfo{person}{Albrecht Schmidt}.}
  \bibinfo{year}{2016}\natexlab{}.
\newblock \showarticletitle{Shifting Gears: User Interfaces in the Age of
  Autonomous Driving}.
\newblock \bibinfo{journal}{\emph{IEEE Pervasive Computing}}
  \bibinfo{volume}{15}, \bibinfo{number}{1} (\bibinfo{year}{2016}),
  \bibinfo{pages}{32--38}.
\newblock


\bibitem[\protect\citeauthoryear{Lauber, B\"{o}ttcher, and Butz}{Lauber
  et~al\mbox{.}}{2014}]%
        {Lauber2014}
\bibfield{author}{\bibinfo{person}{Felix Lauber}, \bibinfo{person}{Claudius
  B\"{o}ttcher}, {and} \bibinfo{person}{Andreas Butz}.}
  \bibinfo{year}{2014}\natexlab{}.
\newblock \showarticletitle{PapAR: Paper Prototyping for Augmented Reality}. In
  \bibinfo{booktitle}{\emph{Adjunct Proceedings of the 6th International
  Conference on Automotive User Interfaces and Interactive Vehicular
  Applications}} (Seattle, WA, USA) \emph{(\bibinfo{series}{AutomotiveUI
  '14})}. \bibinfo{publisher}{Association for Computing Machinery},
  \bibinfo{address}{New York, NY, USA}, \bibinfo{pages}{1–6}.
\newblock
\showISBNx{9781450307253}
\urldef\tempurl%
\url{https://doi.org/10.1145/2667239.2667271}
\showDOI{\tempurl}


\bibitem[\protect\citeauthoryear{Ledo, Houben, Vermeulen, Marquardt, Oehlberg,
  and Greenberg}{Ledo et~al\mbox{.}}{2018}]%
        {Ledo2018}
\bibfield{author}{\bibinfo{person}{David Ledo}, \bibinfo{person}{Steven
  Houben}, \bibinfo{person}{Jo Vermeulen}, \bibinfo{person}{Nicolai Marquardt},
  \bibinfo{person}{Lora Oehlberg}, {and} \bibinfo{person}{Saul Greenberg}.}
  \bibinfo{year}{2018}\natexlab{}.
\newblock \showarticletitle{Evaluation Strategies for HCI Toolkit Research}. In
  \bibinfo{booktitle}{\emph{Proceedings of the 2018 CHI Conference on Human
  Factors in Computing Systems}} (Montreal QC, Canada)
  \emph{(\bibinfo{series}{CHI ’18})}. \bibinfo{publisher}{Association for
  Computing Machinery}, \bibinfo{address}{New York, NY, USA},
  \bibinfo{pages}{1–17}.
\newblock
\showISBNx{9781450356206}
\urldef\tempurl%
\url{https://doi.org/10.1145/3173574.3173610}
\showDOI{\tempurl}


\bibitem[\protect\citeauthoryear{Lee and Kolodge}{Lee and Kolodge}{2018}]%
        {Lee2018}
\bibfield{author}{\bibinfo{person}{John~D. Lee} {and} \bibinfo{person}{Kristin
  Kolodge}.} \bibinfo{year}{2018}\natexlab{}.
\newblock \showarticletitle{Understanding Attitudes Towards Self-Driving
  Vehicles: Quantitative Analysis of Qualitative Data}.
\newblock \bibinfo{journal}{\emph{Proceedings of the Human Factors and
  Ergonomics Society Annual Meeting}} \bibinfo{volume}{62}, \bibinfo{number}{1}
  (\bibinfo{year}{2018}), \bibinfo{pages}{1399--1403}.
\newblock
\urldef\tempurl%
\url{https://doi.org/10.1177/1541931218621319}
\showDOI{\tempurl}
\showeprint{https://doi.org/10.1177/1541931218621319}


\bibitem[\protect\citeauthoryear{Lim, Stolterman, and Tenenberg}{Lim
  et~al\mbox{.}}{2008}]%
        {Lim2008}
\bibfield{author}{\bibinfo{person}{Youn-Kyung Lim}, \bibinfo{person}{Erik
  Stolterman}, {and} \bibinfo{person}{Josh Tenenberg}.}
  \bibinfo{year}{2008}\natexlab{}.
\newblock \showarticletitle{The Anatomy of Prototypes: Prototypes as Filters,
  Prototypes as Manifestations of Design Ideas}.
\newblock \bibinfo{journal}{\emph{ACM Trans. Comput.-Hum. Interact.}}
  \bibinfo{volume}{15}, \bibinfo{number}{2}, Article \bibinfo{articleno}{7}
  (\bibinfo{date}{July} \bibinfo{year}{2008}), \bibinfo{numpages}{27}~pages.
\newblock
\showISSN{1073-0516}
\urldef\tempurl%
\url{https://doi.org/10.1145/1375761.1375762}
\showDOI{\tempurl}


\bibitem[\protect\citeauthoryear{Lockton, Brawley, Ulloa, Prindible, Forlano,
  Rygh, Fass, Herzog, and Nissen}{Lockton et~al\mbox{.}}{2020}]%
        {Lockton2020}
\bibfield{author}{\bibinfo{person}{Dan Lockton}, \bibinfo{person}{Lisa
  Brawley}, \bibinfo{person}{Manuela Ulloa}, \bibinfo{person}{Matt Prindible},
  \bibinfo{person}{Laura Forlano}, \bibinfo{person}{Karianne Rygh},
  \bibinfo{person}{John Fass}, \bibinfo{person}{Katie Herzog}, {and}
  \bibinfo{person}{Bettina Nissen}.} \bibinfo{year}{2020}\natexlab{}.
\newblock \showarticletitle{Tangible Thinking: Materialising how we imagine and
  understand systems, experiences, and relationships}.
\newblock


\bibitem[\protect\citeauthoryear{Macrorie, Marvin, and While}{Macrorie
  et~al\mbox{.}}{2019}]%
        {Macrorie2019}
\bibfield{author}{\bibinfo{person}{Rachel Macrorie}, \bibinfo{person}{Simon
  Marvin}, {and} \bibinfo{person}{Aidan While}.}
  \bibinfo{year}{2019}\natexlab{}.
\newblock \showarticletitle{Robotics and automation in the city: a research
  agenda}.
\newblock \bibinfo{journal}{\emph{Urban Geography}} \bibinfo{volume}{0},
  \bibinfo{number}{0} (\bibinfo{year}{2019}), \bibinfo{pages}{1--21}.
\newblock
\urldef\tempurl%
\url{https://doi.org/10.1080/02723638.2019.1698868}
\showDOI{\tempurl}
\showeprint{https://doi.org/10.1080/02723638.2019.1698868}


\bibitem[\protect\citeauthoryear{Mahadevan, Somanath, and Sharlin}{Mahadevan
  et~al\mbox{.}}{2018}]%
        {Mahadevan2018}
\bibfield{author}{\bibinfo{person}{Karthik Mahadevan}, \bibinfo{person}{Sowmya
  Somanath}, {and} \bibinfo{person}{Ehud Sharlin}.}
  \bibinfo{year}{2018}\natexlab{}.
\newblock \showarticletitle{Communicating Awareness and Intent in Autonomous
  Vehicle-Pedestrian Interaction}. In \bibinfo{booktitle}{\emph{Proceedings of
  the 2018 CHI Conference on Human Factors in Computing Systems}} (Montreal QC,
  Canada) \emph{(\bibinfo{series}{CHI ’18})}. \bibinfo{publisher}{Association
  for Computing Machinery}, \bibinfo{address}{New York, NY, USA},
  \bibinfo{pages}{1–12}.
\newblock
\showISBNx{9781450356206}
\urldef\tempurl%
\url{https://doi.org/10.1145/3173574.3174003}
\showDOI{\tempurl}


\bibitem[\protect\citeauthoryear{Malizia, Chamberlain, and Willcock}{Malizia
  et~al\mbox{.}}{2018}]%
        {Malizia2018}
\bibfield{author}{\bibinfo{person}{Alessio Malizia}, \bibinfo{person}{Alan
  Chamberlain}, {and} \bibinfo{person}{Ian Willcock}.}
  \bibinfo{year}{2018}\natexlab{}.
\newblock \showarticletitle{From Design Fiction to Design Fact: Developing
  Future User Experiences with Proto-Tools}. In
  \bibinfo{booktitle}{\emph{Human-Computer Interaction. Theories, Methods, and
  Human Issues}}, \bibfield{editor}{\bibinfo{person}{Masaaki Kurosu}} (Ed.).
  \bibinfo{publisher}{Springer International Publishing},
  \bibinfo{address}{Cham}, \bibinfo{pages}{159--168}.
\newblock
\showISBNx{978-3-319-91238-7}


\bibitem[\protect\citeauthoryear{Matviienko, Rauschenberger, Cobus, Timmermann,
  M\"{u}ller, Fortmann, L\"{o}cken, Trappe, Heuten, and Boll}{Matviienko
  et~al\mbox{.}}{2015}]%
        {Matviienko2015}
\bibfield{author}{\bibinfo{person}{Andrii Matviienko}, \bibinfo{person}{Maria
  Rauschenberger}, \bibinfo{person}{Vanessa Cobus}, \bibinfo{person}{Janko
  Timmermann}, \bibinfo{person}{Heiko M\"{u}ller}, \bibinfo{person}{Jutta
  Fortmann}, \bibinfo{person}{Andreas L\"{o}cken}, \bibinfo{person}{Christoph
  Trappe}, \bibinfo{person}{Wilko Heuten}, {and} \bibinfo{person}{Susanne
  Boll}.} \bibinfo{year}{2015}\natexlab{}.
\newblock \showarticletitle{Deriving Design Guidelines for Ambient Light
  Systems}. In \bibinfo{booktitle}{\emph{Proceedings of the 14th International
  Conference on Mobile and Ubiquitous Multimedia}} (Linz, Austria)
  \emph{(\bibinfo{series}{MUM ’15})}. \bibinfo{publisher}{Association for
  Computing Machinery}, \bibinfo{address}{New York, NY, USA},
  \bibinfo{pages}{267–277}.
\newblock
\showISBNx{9781450336055}
\urldef\tempurl%
\url{https://doi.org/10.1145/2836041.2836069}
\showDOI{\tempurl}


\bibitem[\protect\citeauthoryear{Nascimento, Muller~Queiroz, Vismari,
  Bailenson, Cugnasca, Junior, and Almeida}{Nascimento et~al\mbox{.}}{2019}]%
        {Nascimento2019}
\bibfield{author}{\bibinfo{person}{Alexandre Nascimento},
  \bibinfo{person}{Anna~Carolina Muller~Queiroz}, \bibinfo{person}{Lucio
  Vismari}, \bibinfo{person}{Jeremy Bailenson}, \bibinfo{person}{Paulo
  Cugnasca}, \bibinfo{person}{Joao Junior}, {and} \bibinfo{person}{Jorge
  Almeida}.} \bibinfo{year}{2019}\natexlab{}.
\newblock \showarticletitle{The Role of Virtual Reality in Autonomous
  Vehicles’ Safety}. \bibinfo{pages}{50--507}.
\newblock
\urldef\tempurl%
\url{https://doi.org/10.1109/AIVR46125.2019.00017}
\showDOI{\tempurl}


\bibitem[\protect\citeauthoryear{Nguyen, Holl\"{a}nder, Hoggenmueller, Parker,
  and Tomitsch}{Nguyen et~al\mbox{.}}{2019}]%
        {Nguyen2019}
\bibfield{author}{\bibinfo{person}{Trung~Thanh Nguyen}, \bibinfo{person}{Kai
  Holl\"{a}nder}, \bibinfo{person}{Marius Hoggenmueller},
  \bibinfo{person}{Callum Parker}, {and} \bibinfo{person}{Martin Tomitsch}.}
  \bibinfo{year}{2019}\natexlab{}.
\newblock \showarticletitle{Designing for Projection-Based Communication
  between Autonomous Vehicles and Pedestrians}. In
  \bibinfo{booktitle}{\emph{Proceedings of the 11th International Conference on
  Automotive User Interfaces and Interactive Vehicular Applications}} (Utrecht,
  Netherlands) \emph{(\bibinfo{series}{AutomotiveUI ’19})}.
  \bibinfo{publisher}{Association for Computing Machinery},
  \bibinfo{address}{New York, NY, USA}, \bibinfo{pages}{284–294}.
\newblock
\showISBNx{9781450368841}
\urldef\tempurl%
\url{https://doi.org/10.1145/3342197.3344543}
\showDOI{\tempurl}


\bibitem[\protect\citeauthoryear{Owensby, Tomitsch, and Parker}{Owensby
  et~al\mbox{.}}{2018}]%
        {Owensby2018}
\bibfield{author}{\bibinfo{person}{Chelsea Owensby}, \bibinfo{person}{Martin
  Tomitsch}, {and} \bibinfo{person}{Callum Parker}.}
  \bibinfo{year}{2018}\natexlab{}.
\newblock \showarticletitle{A Framework for Designing Interactions between
  Pedestrians and Driverless Cars: Insights from a Ride-Sharing Design Study}.
  In \bibinfo{booktitle}{\emph{Proceedings of the 30th Australian Conference on
  Computer-Human Interaction}} (Melbourne, Australia)
  \emph{(\bibinfo{series}{OzCHI ’18})}. \bibinfo{publisher}{Association for
  Computing Machinery}, \bibinfo{address}{New York, NY, USA},
  \bibinfo{pages}{359–363}.
\newblock
\showISBNx{9781450361880}
\urldef\tempurl%
\url{https://doi.org/10.1145/3292147.3292218}
\showDOI{\tempurl}


\bibitem[\protect\citeauthoryear{Pettersson and Ju}{Pettersson and Ju}{2017}]%
        {Pettersson2017}
\bibfield{author}{\bibinfo{person}{Ingrid Pettersson} {and}
  \bibinfo{person}{Wendy Ju}.} \bibinfo{year}{2017}\natexlab{}.
\newblock \showarticletitle{Design Techniques for Exploring Automotive
  Interaction in the Drive towards Automation}. In
  \bibinfo{booktitle}{\emph{Proceedings of the 2017 Conference on Designing
  Interactive Systems}} (Edinburgh, United Kingdom) \emph{(\bibinfo{series}{DIS
  ’17})}. \bibinfo{publisher}{Association for Computing Machinery},
  \bibinfo{address}{New York, NY, USA}, \bibinfo{pages}{147–160}.
\newblock
\showISBNx{9781450349222}
\urldef\tempurl%
\url{https://doi.org/10.1145/3064663.3064666}
\showDOI{\tempurl}


\bibitem[\protect\citeauthoryear{Pfleging, Rang, and Broy}{Pfleging
  et~al\mbox{.}}{2016}]%
        {Pfleging2016}
\bibfield{author}{\bibinfo{person}{Bastian Pfleging}, \bibinfo{person}{Maurice
  Rang}, {and} \bibinfo{person}{Nora Broy}.} \bibinfo{year}{2016}\natexlab{}.
\newblock \showarticletitle{Investigating User Needs for Non-Driving-Related
  Activities during Automated Driving}. In
  \bibinfo{booktitle}{\emph{Proceedings of the 15th International Conference on
  Mobile and Ubiquitous Multimedia}} (Rovaniemi, Finland)
  \emph{(\bibinfo{series}{MUM ’16})}. \bibinfo{publisher}{Association for
  Computing Machinery}, \bibinfo{address}{New York, NY, USA},
  \bibinfo{pages}{91–99}.
\newblock
\showISBNx{9781450348607}
\urldef\tempurl%
\url{https://doi.org/10.1145/3012709.3012735}
\showDOI{\tempurl}


\bibitem[\protect\citeauthoryear{Philipsen, Brell, and Ziefle}{Philipsen
  et~al\mbox{.}}{2019}]%
        {Philipsen2019}
\bibfield{author}{\bibinfo{person}{Ralf Philipsen}, \bibinfo{person}{Teresa
  Brell}, {and} \bibinfo{person}{Martina Ziefle}.}
  \bibinfo{year}{2019}\natexlab{}.
\newblock \showarticletitle{Carriage Without a Driver -- User Requirements for
  Intelligent Autonomous Mobility Services}. In
  \bibinfo{booktitle}{\emph{Advances in Human Aspects of Transportation}},
  \bibfield{editor}{\bibinfo{person}{Neville Stanton}} (Ed.).
  \bibinfo{publisher}{Springer International Publishing},
  \bibinfo{address}{Cham}, \bibinfo{pages}{339--350}.
\newblock
\showISBNx{978-3-319-93885-1}


\bibitem[\protect\citeauthoryear{Radianti, Majchrzak, Fromm, and
  Wohlgenannt}{Radianti et~al\mbox{.}}{2020}]%
        {Radianti2020}
\bibfield{author}{\bibinfo{person}{Jaziar Radianti}, \bibinfo{person}{Tim~A.
  Majchrzak}, \bibinfo{person}{Jennifer Fromm}, {and} \bibinfo{person}{Isabell
  Wohlgenannt}.} \bibinfo{year}{2020}\natexlab{}.
\newblock \showarticletitle{A systematic review of immersive virtual reality
  applications for higher education: Design elements, lessons learned, and
  research agenda}.
\newblock \bibinfo{journal}{\emph{Computers \& Education}}
  \bibinfo{volume}{147} (\bibinfo{year}{2020}), \bibinfo{pages}{103778}.
\newblock
\showISSN{0360-1315}
\urldef\tempurl%
\url{https://doi.org/10.1016/j.compedu.2019.103778}
\showDOI{\tempurl}


\bibitem[\protect\citeauthoryear{Rouchitsas and Alm}{Rouchitsas and
  Alm}{2019}]%
        {Rouchitsas2019}
\bibfield{author}{\bibinfo{person}{Alexandros Rouchitsas} {and}
  \bibinfo{person}{Håkan Alm}.} \bibinfo{year}{2019}\natexlab{}.
\newblock \showarticletitle{External Human–Machine Interfaces for Autonomous
  Vehicle-to-Pedestrian Communication: A Review of Empirical Work}.
\newblock \bibinfo{journal}{\emph{Frontiers in Psychology}}
  \bibinfo{volume}{10} (\bibinfo{year}{2019}), \bibinfo{pages}{2757}.
\newblock
\showISSN{1664-1078}
\urldef\tempurl%
\url{https://doi.org/10.3389/fpsyg.2019.02757}
\showDOI{\tempurl}


\bibitem[\protect\citeauthoryear{Rygh and Clatworthy}{Rygh and
  Clatworthy}{2019}]%
        {Rygh2019}
\bibfield{author}{\bibinfo{person}{Karianne Rygh} {and} \bibinfo{person}{Simon
  Clatworthy}.} \bibinfo{year}{2019}\natexlab{}.
\newblock \bibinfo{booktitle}{\emph{The Use of Tangible Tools as a Means to
  Support Co-design During Service Design Innovation Projects in Healthcare}}.
\newblock \bibinfo{publisher}{Springer International Publishing},
  \bibinfo{address}{Cham}, \bibinfo{pages}{93--115}.
\newblock
\showISBNx{978-3-030-00749-2}
\urldef\tempurl%
\url{https://doi.org/10.1007/978-3-030-00749-2_7}
\showDOI{\tempurl}


\bibitem[\protect\citeauthoryear{Sanders, Brandt, and Binder}{Sanders
  et~al\mbox{.}}{2010}]%
        {Sanders2010}
\bibfield{author}{\bibinfo{person}{Elizabeth B.-N. Sanders},
  \bibinfo{person}{Eva Brandt}, {and} \bibinfo{person}{Thomas Binder}.}
  \bibinfo{year}{2010}\natexlab{}.
\newblock \showarticletitle{A Framework for Organizing the Tools and Techniques
  of Participatory Design}. In \bibinfo{booktitle}{\emph{Proceedings of the
  11th Biennial Participatory Design Conference}} (Sydney, Australia)
  \emph{(\bibinfo{series}{PDC ’10})}. \bibinfo{publisher}{Association for
  Computing Machinery}, \bibinfo{address}{New York, NY, USA},
  \bibinfo{pages}{195–198}.
\newblock
\showISBNx{9781450301312}
\urldef\tempurl%
\url{https://doi.org/10.1145/1900441.1900476}
\showDOI{\tempurl}


\bibitem[\protect\citeauthoryear{Schmidt}{Schmidt}{2017}]%
        {Schmidt2017}
\bibfield{author}{\bibinfo{person}{Albrecht Schmidt}.}
  \bibinfo{year}{2017}\natexlab{}.
\newblock \showarticletitle{Understanding and Researching through Making: A
  Plea for Functional Prototypes}.
\newblock \bibinfo{journal}{\emph{Interactions}} \bibinfo{volume}{24},
  \bibinfo{number}{3} (\bibinfo{date}{April} \bibinfo{year}{2017}),
  \bibinfo{pages}{78–81}.
\newblock
\showISSN{1072-5520}
\urldef\tempurl%
\url{https://doi.org/10.1145/3058498}
\showDOI{\tempurl}


\bibitem[\protect\citeauthoryear{Schroeter, Rakotonirainy, and Foth}{Schroeter
  et~al\mbox{.}}{2012}]%
        {Schroeter2012}
\bibfield{author}{\bibinfo{person}{Ronald Schroeter}, \bibinfo{person}{Andry
  Rakotonirainy}, {and} \bibinfo{person}{Marcus Foth}.}
  \bibinfo{year}{2012}\natexlab{}.
\newblock \showarticletitle{The social car: new interactive vehicular
  applications derived from social media and urban informatics}.
\newblock In \bibinfo{booktitle}{\emph{Automotive UI '12: Proceedings of the
  4th International Conference on Automotive User Interfaces and Interactive
  Vehicular Applications}}, \bibfield{editor}{\bibinfo{person}{A~L Kun},
  \bibinfo{person}{L~N Boyle}, \bibinfo{person}{B~Reimer}, {and}
  \bibinfo{person}{A~Riener}} (Eds.). \bibinfo{publisher}{Association for
  Computing Machinery}, \bibinfo{address}{United States},
  \bibinfo{pages}{107--110}.
\newblock
\urldef\tempurl%
\url{https://doi.org/10.1145/2390256.2390273}
\showDOI{\tempurl}


\bibitem[\protect\citeauthoryear{Stadler, Cornet, Novaes~Theoto, and
  Frenkler}{Stadler et~al\mbox{.}}{2019}]%
        {Stadler2019}
\bibfield{author}{\bibinfo{person}{Sebastian Stadler},
  \bibinfo{person}{Henriette Cornet}, \bibinfo{person}{Tatiana Novaes~Theoto},
  {and} \bibinfo{person}{Fritz Frenkler}.} \bibinfo{year}{2019}\natexlab{}.
\newblock \bibinfo{booktitle}{\emph{A Tool, not a Toy: Using Virtual Reality to
  Evaluate the Communication Between Autonomous Vehicles and Pedestrians}}.
\newblock \bibinfo{publisher}{Springer International Publishing},
  \bibinfo{address}{Cham}, \bibinfo{pages}{203--216}.
\newblock
\showISBNx{978-3-030-06246-0}
\urldef\tempurl%
\url{https://doi.org/10.1007/978-3-030-06246-0_15}
\showDOI{\tempurl}


\bibitem[\protect\citeauthoryear{Tomitsch and Hoggenmueller}{Tomitsch and
  Hoggenmueller}{2020}]%
        {Tomitsch2020}
\bibfield{author}{\bibinfo{person}{Martin Tomitsch} {and}
  \bibinfo{person}{Marius Hoggenmueller}.} \bibinfo{year}{2020}\natexlab{}.
\newblock \showarticletitle{Designing Human-Machine Interactions in the
  Automated City: Methodologies, Considerations, Principles}. In
  \bibinfo{booktitle}{\emph{Automating Cities: Robots, Drones and Urban Data in
  the Design, Construction, Operation and Future Impact of Smart Cities}},
  \bibfield{editor}{\bibinfo{person}{Chien Ming~Wang} {and}
  \bibinfo{person}{Brydon~Timothy Wang}} (Eds.). \bibinfo{publisher}{Springer}.
\newblock


\bibitem[\protect\citeauthoryear{Wiethoff and Bl\"{o}ckner}{Wiethoff and
  Bl\"{o}ckner}{2010}]%
        {Wiethoff2010}
\bibfield{author}{\bibinfo{person}{Alexander Wiethoff} {and}
  \bibinfo{person}{Magdalena Bl\"{o}ckner}.} \bibinfo{year}{2010}\natexlab{}.
\newblock \showarticletitle{LightBox: Exploring Interaction Modalities with
  Colored Light}. In \bibinfo{booktitle}{\emph{Proceedings of the Fifth
  International Conference on Tangible, Embedded, and Embodied Interaction}}
  (Funchal, Portugal) \emph{(\bibinfo{series}{TEI ’11})}.
  \bibinfo{publisher}{Association for Computing Machinery},
  \bibinfo{address}{New York, NY, USA}, \bibinfo{pages}{399–400}.
\newblock
\showISBNx{9781450304788}
\urldef\tempurl%
\url{https://doi.org/10.1145/1935701.1935799}
\showDOI{\tempurl}


\bibitem[\protect\citeauthoryear{Wiethoff, Schneider, K\"{u}fner, Rohs, Butz,
  and Greenberg}{Wiethoff et~al\mbox{.}}{2013}]%
        {Wiethoff2013}
\bibfield{author}{\bibinfo{person}{Alexander Wiethoff}, \bibinfo{person}{Hanna
  Schneider}, \bibinfo{person}{Julia K\"{u}fner}, \bibinfo{person}{Michael
  Rohs}, \bibinfo{person}{Andreas Butz}, {and} \bibinfo{person}{Saul
  Greenberg}.} \bibinfo{year}{2013}\natexlab{}.
\newblock \showarticletitle{Paperbox: A Toolkit for Exploring Tangible
  Interaction on Interactive Surfaces}. In
  \bibinfo{booktitle}{\emph{Proceedings of the 9th ACM Conference on Creativity
  \& Cognition}} (Sydney, Australia) \emph{(\bibinfo{series}{C\&C '13})}.
  \bibinfo{publisher}{Association for Computing Machinery},
  \bibinfo{address}{New York, NY, USA}, \bibinfo{pages}{64–73}.
\newblock
\showISBNx{9781450321501}
\urldef\tempurl%
\url{https://doi.org/10.1145/2466627.2466635}
\showDOI{\tempurl}


\bibitem[\protect\citeauthoryear{Wiethoff, Schneider, Rohs, Butz, and
  Greenberg}{Wiethoff et~al\mbox{.}}{2012}]%
        {Wiethoff2012}
\bibfield{author}{\bibinfo{person}{Alexander Wiethoff}, \bibinfo{person}{Hanna
  Schneider}, \bibinfo{person}{Michael Rohs}, \bibinfo{person}{Andreas Butz},
  {and} \bibinfo{person}{Saul Greenberg}.} \bibinfo{year}{2012}\natexlab{}.
\newblock \showarticletitle{Sketch-a-TUI: Low Cost Prototyping of Tangible
  Interactions Using Cardboard and Conductive Ink}. In
  \bibinfo{booktitle}{\emph{Proceedings of the Sixth International Conference
  on Tangible, Embedded and Embodied Interaction}} (Kingston, Ontario, Canada)
  \emph{(\bibinfo{series}{TEI ’12})}. \bibinfo{publisher}{Association for
  Computing Machinery}, \bibinfo{address}{New York, NY, USA},
  \bibinfo{pages}{309–312}.
\newblock
\showISBNx{9781450311748}
\urldef\tempurl%
\url{https://doi.org/10.1145/2148131.2148196}
\showDOI{\tempurl}


\end{thebibliography}
\end{document}